# The heterogeneous near-surface velocity structure of a carbonate-hosted seismogenic fault zone and its dependence on the investigated length scale

Michele Fondriest[1,2*], Thomas M. Mitchell[3], Maurizio Vassallo[4], Stephane Garambois[2], Giuseppe Di Giulio[4], Fabrizio Balsamo[5], Marta Pischiutta[4], Mai-Linh Doan[2]

[1] Dipartimento di Geoscienze, Università degli Studi di Padova, Padua, Italy
[2] Institut des Sciences de la Terre (ISTerre), Université Grenoble Alpes, Grenoble, France
[3] Department of Earth Sciences, University College of London, London, United Kingdom
[4] Istituto Nazionale di Geofisica e Vulcanologia (INGV), Rome and L'Aquila, Italy
[5] Dipartimento di Scienze Chimiche, della Vita e della Sostenibilità Ambientale, Università di Parma, Parma, Italy

*corresponding author: Michele Fondriest, michele.fondriest@unipd.it

## Abstract

Field geological studies highlighted the heterogeneous structure of fault zones from the meter- to millimeter scale, but such internal variability is not generally resolved by seismological techniques due to spatial resolution limits. The near-surface velocity structure of the Vado di Corno seismogenic fault zone was quantified at different length scales, from laboratory measurements of ultrasonic velocities (few centimeters rock samples, 1 MHz source) to high-resolution first-arrival seismic tomography (spatial resolution to a few meters). The fault zone juxtaposed structural units with contrasting ultrasonic velocities. The fault core cataclastic units were "slower" compared to damage zone units. A negative correlation between ultrasonic velocity and porosity was observed, with dispersion in fault core units related to varying degree of textural maturity and pore space sealing by calcite. Low-velocity outliers in the damage zone were instead linked to microfracture networks with local cataclasis and partial calcite sealing. P-wave high-resolution seismic tomography imaged distinct fault-bounded rock bodies, matching the geometry and size of field-mapped structural units. At this length scale, relatively "fast" fault core units and low-strain damage zones contrasted with a very "slow" intensely fractured high-strain damage zone. The discrepancy between higher ultrasonic velocities and lower tomography-derived ones was reconciled through an effective medium approach considering the effect of meso-scale fractures in each unit. This revealed a heterogeneous fault zone velocity structure with different scaling among structural units. Lastly, the persistence of a thick compliant high-strain damage zone at shallow depth may significantly affect fault zone mechanics and the distribution of near-surface deformations.

## Plain language summary

The internal structure of fault zones controls their mechanical and hydrologic behavior. Geological observations highlight the heterogeneous structure of fault zones, characterized by rock units with distinct deformation and physical properties, but such internal variability is often not detectable within the low-velocity zones imaged by conventional seismic surveys. Therefore, determining wave velocities and related physical properties in fault zones at different observational scales becomes crucial. For this purpose, the near-surface velocity structure of an exhumed active fault zone was quantified at different length scales, from the laboratory on few centimeters in size rock samples to the field through a specifically tailored high-resolution seismic survey. In the laboratory, the fault core units were "slower" than the damage zone units and the ultrasonic velocities were controlled by fault rocks porosity and microstructure. The high-resolution seismic survey imaged fault-bounded rock bodies matching with the rock units mapped in the field. At this scale "fast" fault core units and low-strain damage zone contrasted with a very “slow” high-strain damage zone. The discrepancy between laboratory- and field-scale velocities was reconciled by quantifying the effect of mesoscale fractures for each rock unit. The implications of this heterogenous velocity structure for near-surface fault zone mechanics are discussed.

**Key points**

- The Vado di Corno fault zone juxtaposes structural units with strong ultrasonic and seismic velocities contrasts.
- Fault core ultrasonic velocities are lower than damage zone ones and are strongly influenced by textural maturity and pore space sealing.
- At the seismic scale, stiff fault core and low-strain damage zone contrast with very compliant and fractured high-strain damage zone.

## 1. Introduction

Fault zones in the upper crust are characterized by a heterogenous internal structure arising from the juxtaposition of rock volumes with distinct deformation intensity and physical-transport properties (Caine et al., 1996; Chester et al., 1993; Faulkner et al., 2010; Wibberley et al., 2008). This spatial variability may be controlled by lateral contrasts in lithology, density of faults and fractures, mineralogy, porosity and microstructures over length scales in the range $10^{-3}$-$10^{2}$ m (Schulz and Evans, 1998, 2000). Furthermore, the magnitude and orientation of local and regional stresses together with pore pressure and temperature levels, determine non-linear changes with depth in the mechanical behavior of rocks and fluid flow within fault zones (Evans et al., 1997; Scholz, 2012; Zoback and Townend, 2001). A range of geological observations from both active and exhumed fault zones show that most of the fault slip is accommodated within a fault core few meters thick comprising layers of low-permeability cataclasites and gouges with evidence of extreme shear strain localization along principal slip zones (Sibson et al., 2003; Smith et al., 2011). The adjacent damage zone is instead represented by a 100s meters thick, fluid-permeable and low-shear strain unit with macro- and microfractures density progressively decreasing to a background level moving toward the less deformed country rocks (Caine et al., 1996; Choi et al., 2016; Kim et al., 2004; Mitchell et al., 2009). Further structural complexities and heterogeneity may develop due to significant slip variability along fault strike and dip, to the interaction of segmented fault networks, and to the effect of inherited structures (e.g., Childs et al., 2009; Cowie and Scholz, 1992; Fondriest et al., 2012, 2020; Masoch et al., 2021, 2022; Mitchell et al., 2009; Myers and Aydin, 2004; Ostermeijer et al., 2020).

The elastic properties of fault zone rocks, i.e. their elastic moduli and wave speeds, commonly show reduced values compared to the relatively undeformed country rocks, due to an increase in fracturing and/or alteration (e.g. Stierman, 1984). Depending on their magnitude, spatial extent and distribution, these reductions can significantly affect the mechanics of fault zones during the seismic cycle (Faulkner et al., 2006 and 2010). For instance, the occurrence of large compliant damage zones can potentially modify the dynamics of seismic ruptures by modulating their propagation speed, the rupture style and the radiated energy through multiple feedback mechanisms (e.g. Huang et al., 2014; Xu et al., 2014). On the other hand, the heterogenous distribution of elastic properties near the Earth's surface can affect the amplification of ground motion and the distribution of both long-term and co-seismic shallow deformations (Duan at al., 2011; Catchings et al., 2014; Nevitt et al., 2021). Given this wide range of possible interactions, it is of paramount importance to quantify elastic properties over different length scales within fault zones.

Laboratory measurements of elastic wave speeds in rocks are mostly performed at length scales of few centimeters and at ultrasonic frequencies of 0.1-2 MHz which are significantly larger than those of surface seismic surveys (~ 3-300 Hz; Birch, 1960). The P- and S- ultrasonic wave velocities of undeformed country rocks with different composition have been extensively studied since 1960s and empirical relationships between velocity and intrinsic (porosity, microstructure, mineralogy) and extrinsic (effective pressure, temperature) factors have been determined (e.g. Anselmetti and Eberli, 1993; Castagna et al., 1985; Marion et al., 1992). In exploration geology, these data are often combined with sonic log velocities, usually acquired at 10s kHz and length scales of few meters, to infer the physical properties of reservoir rocks and to calibrate seismic reflection data (Biddle et al., 1992; Campbell and Stafleu, 1992; Christensen and Szymanski, 1991; Sellami et al., 1990). More recently, direct measurements of ultrasonic waves have been performed on fault zone rocks in crystalline and sedimentary rocks, both siliciclastic and carbonates (Adam et al., 2020; Allen et al., 2017; Boulton et al., 2012; Carpenter et al., 2014; Ferraro et al., 2020; Jeppson and Tobin, 2020; Rempe et al., 2013, 2018; Trippetta et al., 2017). These studies aimed to quantify the relationships between seismic velocity and porosity type/distribution, microfracture density, or permeability in rock samples representative of fault zone structural units with distinct deformation intensity (Agosta et al., 2007; Matonti et al., 2012, 2015). However, the proposed elastic models are mostly based on the interpolation of such short-scale velocity measurements over the entire fault zones, thus neglecting the strong effect of meso- to macroscale heterogeneities (e.g. open fractures) on the elastic wave speeds (see Jeanne et al., 2012; Matonti et al., 2015, 2017; Stierman and Kovach, 1979 about this point). Beyond this, little is still known about the dispersion of rocks elastic properties measured at the laboratory scale over a wide range of frequencies, from seismic to ultrasonic (3-3000 Hz; Adam et al., 2006, 2008; Pimienta et al., 2015).
The geophysical expression of fault zones by seismic refraction methods in a broad sense is represented by regions of low seismic wave velocities due to the reduction in the effective elastic moduli of rocks (Mooney et al., 1986; Ben-Zion and Sammis, 2003). At the crustal scale, both active (large borehole explosions; e.g. Li et al., 2004) and passive surveys (earthquakes or seismic noise recorded at regional arrays) have recognized lateral and vertical changes in surface and body wave velocities (10%-50% reduction) at fault zones up to several kilometers deep, along with the characteristic appearance of trapped and head fault zone waves (Hillers et al., 2016; Rovelli et al., 2002). However, the spatial resolution of these surveys is low and typically ranges from hundreds of meters to tens of kilometers. On the other hand, ambient noise cross-correlation on dense arrays, and even more so, high-resolution active seismic surveys, have the potential to retrieve fine-scale images of the shallow velocity structure of fault zones with a spatial resolution of several meters which is comparable to the scale of heterogeneity described by field geology (e.g. Roux et al. 2016). Nevertheless, many near-surface seismic surveys are used to identify faults not evident at the surface and are limited in depth to the investigation of Quaternary deposits covering the deformed bedrock units (Catchings et al. 2002).
In this study, we quantified the near-surface velocity structure of a well-exposed section of an extensional seismogenic fault zone (Vado di Corno Fault Zone, Central Apennines, Italy) at different length and frequency scales, from laboratory ultrasonic velocity measurements to a high-resolution active field seismic survey. The fault zone juxtaposes Quaternary deposits in the hanging wall with a heterogeneous footwall block consisting of variably deformed carbonates exhumed from depth. Based on the detailed mapping by Fondriest et al. (2020), each distinct structural unit was systematically across the fault zone footwall and characterized in terms of directional ultrasonic wave speeds (P- and S-waves), porosity, mineralogy and microstructures. A high-resolution P-wave seismic survey was performed along a profile laterally overlapping with the sampling transect, to obtain a first-arrival tomography image of the fault zone. This combined study allowed us (i) to investigate the factors controlling elastic wave velocities from the ultrasonic to the seismic length and frequency scales, (ii) to quantify for the first time at high-resolution the near-surface seismic velocity structure of an exhumed carbonate-hosted seismogenic source of the region, and (iii) to highlight the strong heterogeneity and scale-dependency of the fault zone elastic properties in contrast with the homogeneous low-velocity zones detected by larger-scale seismic surveys. These new findings have strong implications for the mechanical behavior of fault zones in the near-surface during the seismic cycle.

## 2. The internal structure of the Vado di Corno Fault Zone (VCFZ)

The Vado di Corno Fault Zone (VCFZ) is a segment of the seismically active Campo Imperatore extensional fault system which strikes WNE-ESE for ~20 km in the Gran Sasso Range (Central Apennines, Italy) and has been associated by paleoseismological studies with three surface ruptures with potential for $M_W$ 7 earthquakes since late Holocene (Galadini et al, 2003; Galli et al., 2002, 2022) (Fig.1a). The VCFZ borders to the north the upper part of the Campo Imperatore intramountain basin (Vezzani et al., 2010) and is characterized by a thick continuous band of intensely fractured carbonates beautifully exposed in its footwall block exhumed from depths down to ~2 km (Agosta and Kirschner, 2003). The exposed hanging wall block, on the other hand, consists of glacial, alluvial, and colluvial deposits of Pleistocene to Holocene age (Adamoli et al., 2012; Giraudi and Frezzotti, 1995), implying a total extensional slip of 1.5–2 km across the fault zone (Ghisetti and Vezzani, 1991).
Here, we build on a previous study that quantified in detail the 3D fault network and fault zone rock distribution in the VCFZ footwall block along a segment of ~2 km (Fondriest et al., 2020; Fig.1b). The VCFZ consists of an array of extensional fault surfaces with dominant WNW-ESE strike, mostly dipping to the SSW, and cuts through the inherited Omo Morto Thrust Zone (Pliocene age) with partial reactivation in extension (Figs.1b-2; Demurtas et al., 2016; Fondriest et al., 2020). Antithetic and transverse extensional faults occur in the external portion of the fault zone (Fig.1b).
The stratigraphic units affected by the VCFZ are the dolomitized Calcare Massiccio, lying in a SW dipping monocline on top of the Omo Morto thrust, and some strongly deformed beds of Verde Ammonitico outcropping in tectonic windows (Demurtas et al., 2016). The along-strike segmentation and strain distribution within the VCFZ were strongly controlled by the complex original geometry of the inherited thrust zone (Fondriest et al., 2020). In particular, five structural units were defined and mapped within the VCFZ, each characterized by a different degree of fracturing and/or cataclastic deformation recognizable at the mesoscale (Demurtas et al., 2016; Fondriest et al., 2020; Figs. 1-4). The structural units, moving from the principal fault surface into the footwall, are: Cataclastic Unit 1 (CU1), Cataclastic Unit 2 (CU2), Breccia Unit (BU), High Strain Damage Zone (HSDZ), Low Strain Damage Zone (LSDZ). Table 1 gathers the key mesoscale features of each unit.
More specifically, CU1 and CU2 form the fault core of the extensional VCFZ and truncate unit BU, which outcrops within the extensional damage zone and represents the inherited Omo Morto Thrust Zone (Figs.2-3). Fault rocks of CU1 and CU2 pertain to the "cataclastic series" (Sibson, 1977), are cemented by calcite and bounded or crosscut by cohesive and highly localized slipping zones (<1–2 cm thick) (Fig.3-4a). Unit CU1 represents the proximal fault core (whitish cataclasites to ultracataclasites) accommodating more shear deformation compared to unit CU2 formed by less mature fault rocks (gray to brownish cataclasites to crush microbreccias; Fig.3-4a; Demurtas et al., 2016). Unit BU consists of crush breccias (*sensu* Sibson, 1977) affected by secondary dolomitization and dolomite veins up to several millimeters thick (Fig.3-4b; Demurtas et al., 2016). The extensional damage zone units HSDZ and LSDZ consist of fractured volumes of dolomitized Calcare Massiccio affected by subsidiary faults, joint sets, thin calcite veins and thicker dolomite veins, the latter associated with the older compressional phase. The HSDZ is characterized by an average fracture spacing of few centimeters which decreases around subsidiary faults (Table 1; Figs.4c-d), while the LSDZ has an average fracture spacing of tens of centimeters and subsidiary faults are spaced few meters apart (Table 1; Figs.4e-f). Here, with the term fracture we mean joints (mode I fractures), dominant in the HSDZ and LSDZ, and few mesoscale shear fractures with sub-centimetric slip. The transition between the HSDZ and the LSDZ occurs by moving away from the Vado di Corno principal fault surface, even if few isolated lenses of LSDZ are embedded within the HSDZ at smaller distance from the fault core (Figs.1b, 2). The current study focuses on the same fault zone section described by Demurtas et al. (2016) (Fig.2), where both the sampling for laboratory measurements of ultrasonic velocities (Figs.2a-3) and the high-resolution seismic survey (profile t-t'; Fig.2a) were performed.

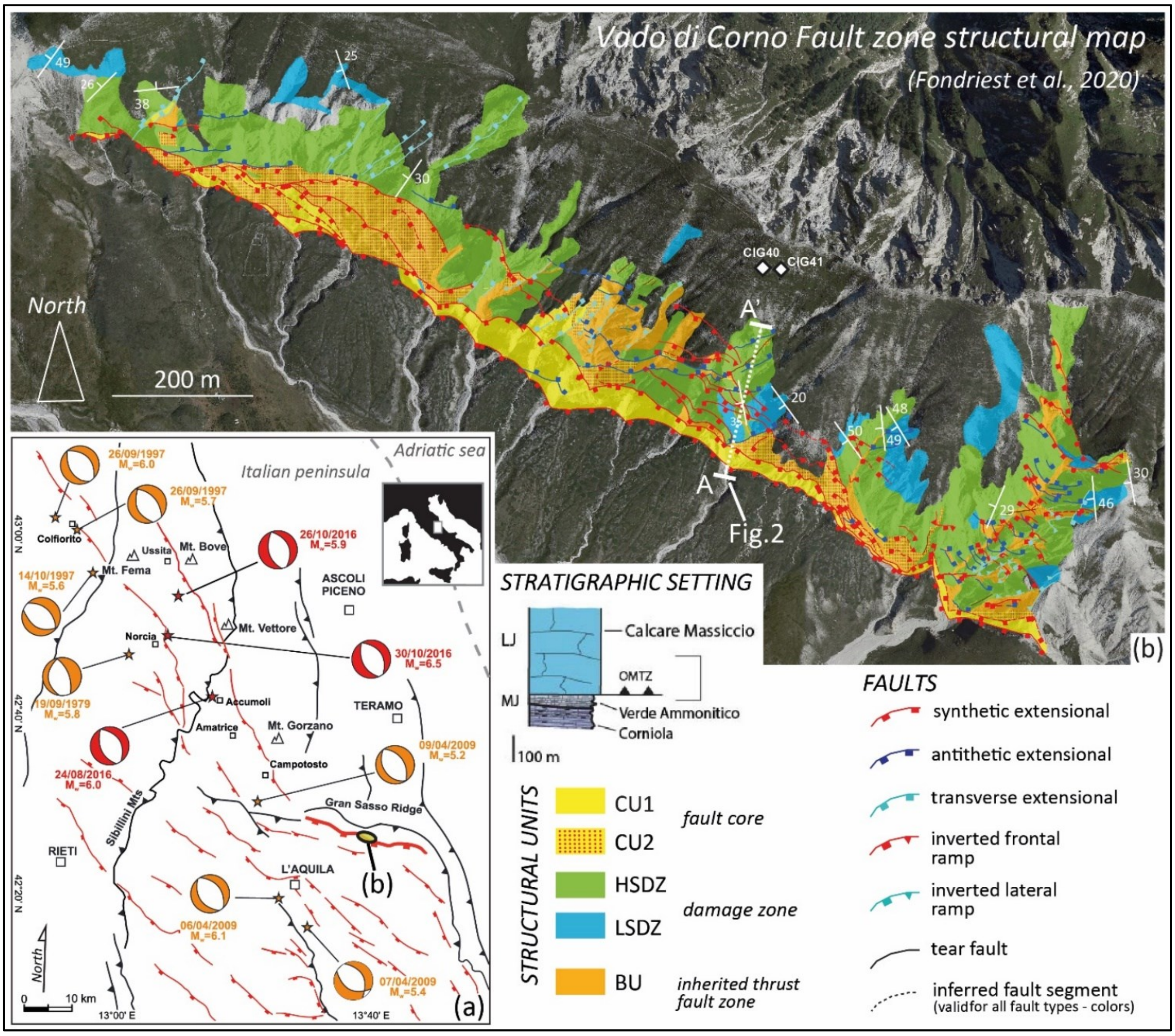

**Fig.1** (a) Seismotectonic map of the northern portion of the Central Apennines and the southern portion of the Northern Apennines of Italy (modified after Fondriest et al., 2020). Main thrust faults are in black and Quaternary active extensional faults are in red. The thicker red line indicates the Campo Imperatore Fault system, part of which is investigated in this study. Focal mechanisms indicate the mainshocks of the seismic sequences which stroke the region in the last 40 years (in red the mainshocks of the 2016 seismic sequence, in orange all the others). (b) Interpreted structural map of the Vado di Corno Fault Zone (VCFZ) footwall block after Fondriest et al. (2020). Distinct structural units are distinguished with different colors: fault core cataclastic units (CU1) and (CU2) in yellow and red dotted yellow, respectively; high-strain damage zone (HSDZ) in green; low-strain damage zone (LSDZ) in blue; and breccia unit (BU) in orange. The BU represents an older thrust fault zone (OMTZ) (Ghisetti and Vezzani, 1991). Principal and subsidiary faults are distinguished as synthetic extensional, antithetic extensional, transverse extensional, inverted frontal ramps, inverted lateral ramps, tear faults, and inferred fault segments. Bedding attitude data are in white symbols. In the VCFZ footwall block the Calcare Massiccio Fm. (Lower Jurassic) is thrusted over Verde Ammonitico Fm. (Middle Jurassic) (see inset with tectonostratigraphic scheme). The trace of section A-A' (Fig.2) and the location of samples CIG40-41 are also reported.

| Structural units | CU1 | CU2 | BU | HSDZ | LSDZ |
|---|---|---|---|---|---|
| Fault zone rock *(Sibon, 1977)* | ultracataclasite to cataclasite | cataclasite to crush microbreccia | crush breccia | fractured rock mass | fractured rock mass |
| Host rock mineralogy | dolomite | dolomite | dolomite | dolomite | dolomite |
| Clast framework mineralogy | dolomite | dolomite | dolomite | - | - |
| Cement | secondary calcite (≤50%) | secondary calcite | secondary dolomite | - | - |
| Veins | diffuse calcite veinlets | diffuse calcite veinlets | dolomite veins - thin calcite veins | dolomite veins - thin calcite veins | dolomite veins - thin calcite veins |
| Average fault spacing | 0.1-0.2 m | 0.1-0.2 m | 2-5 m | 2-3 m | 4-5 m |
| Average fracture spacing | irregular: few joints | irregular: few joints | 0.03-0.4 m | 0.03-0.1 m | 0.3-0.4 m |

**Table 1.** Summary of the mesoscale key characteristics of the VCFZ structural units.

## 3. Materials and Methods

### 3.1. Collection and preparation of oriented samples

Twenty-nine blocks of fault zone rocks (Figs.1b, 2a) were extracted from the VCFZ footwall at increasing distance from the principal fault surface along a profile ~200 m long (Table S1). The blocks had an average size of 40 by 40 cm, were collected spatially oriented and avoiding any mesoscale fault surface. This was particularly important since the VCFZ is characterized by abrupt changes of deformation intensity due to the juxtaposition by faults of distinct structural units (Figs.1-2), and allowed us to track bulk variations of physical properties potentially detectable by the high-resolution seismic survey (see section 3.5.). Since some friable fault rocks were difficult to sample, a high-viscosity two-components resin was poured on specific rock outcrops to impregnate their surface and subsequently carve the blocks.

Back in the laboratory, the blocks were reoriented and fixed in position by casting a base layer of plaster (Fig.5), before being cut and rectified into rectangular parallelepipeds (n=29) with sides of 3-6 cm. Opposite sample faces were ground flat and parallel using a surface grinder. Two faces per sample were cut parallel to the attitude of the VCFZ principal fault surface or to the most nearby high-hierarchy subsidiary fault (the faults reported in the map and section of Fig.2 and Table S1). This allowed us to measure ultrasonic velocities on the same specimen in three mutual orthogonal directions: fault perpendicular, fault parallel horizontal and fault parallel vertical. Additional rock slabs and cylinders (n=22) were prepared with two parallel bases orthogonal to the vertical direction (Table S1). The dimensions of the rectified samples were measured with a digital caliper with associated errors on lengths of ±0.02-0.2 mm. The final rectified samples did not contain open visible fractures. This implies that the samples of HSDZ and LSDZ were representative of the less deformed rocks in between the mesoscale fracture sets (i.e. joints) present in the field exposures.

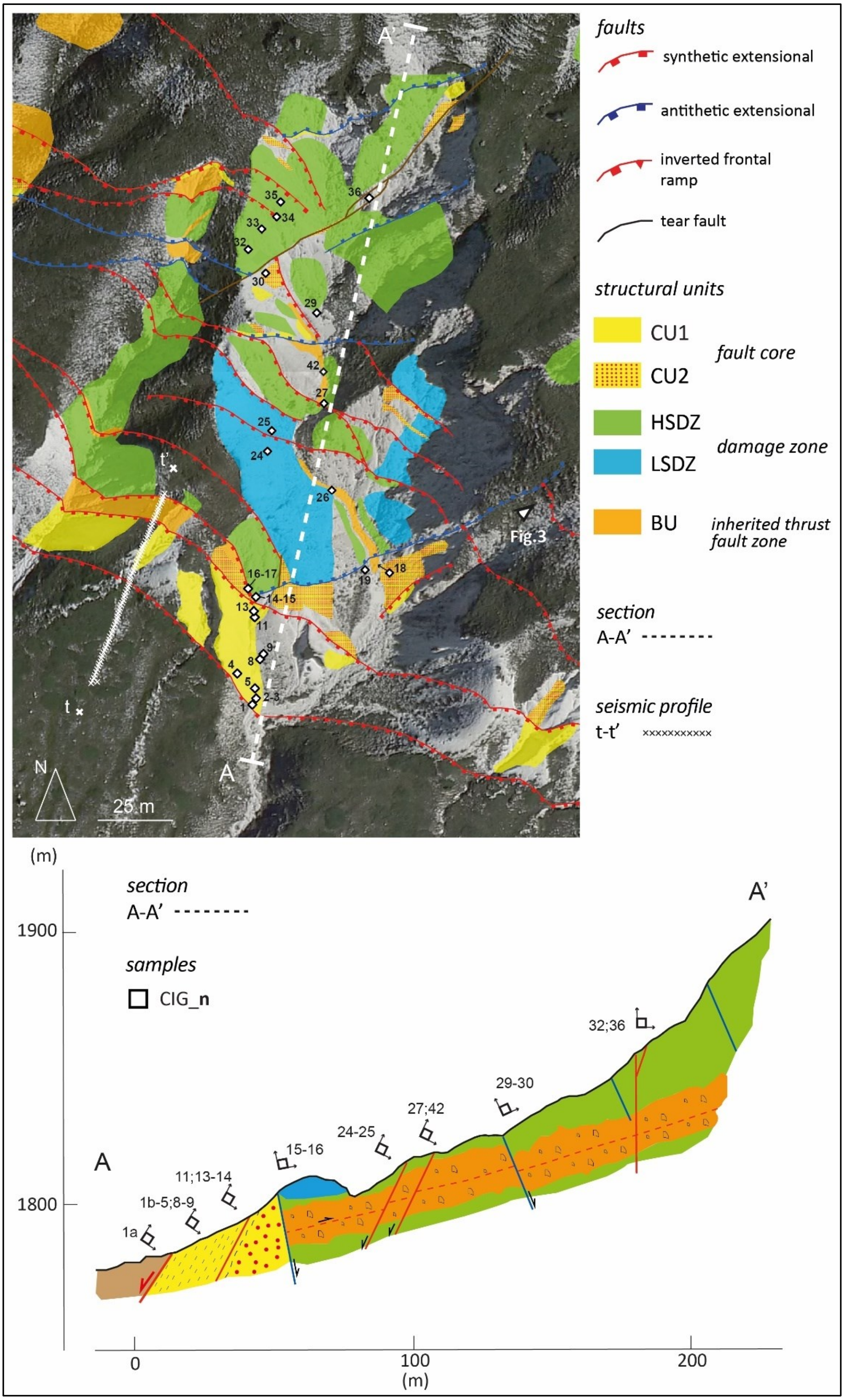

**Fig.2** Outcrop-based structural map of the VCFZ around the section A-A' (left side). The legend of the map is the same of Fig.1. The location of the rock samples used for ultrasonic velocity and petrophysical measurements is indicated across the fault zone (sample number is n, to complete sample name as CIG n; see Table S1). Structural section of the VCFZ footwall along the trace A-A' (right side; see Fondriest et al., 2020). The fault core structural units in yellow (CU1 and CU2) cut the extensional damage zone comprising high- and low-strain damage zones (HSDZ in green, LSDZ in light blue). The breccia unit (BU) is recognized within the extensional damage zone and represented an older thrust zone. The position of the samples is projected, together with their orientation with respect to the principal fault surface and the major subsidiary faults. The trace of the active seismic profiles (t-t') is reported (white crosses are the positions of the GPS points taken along the profile).

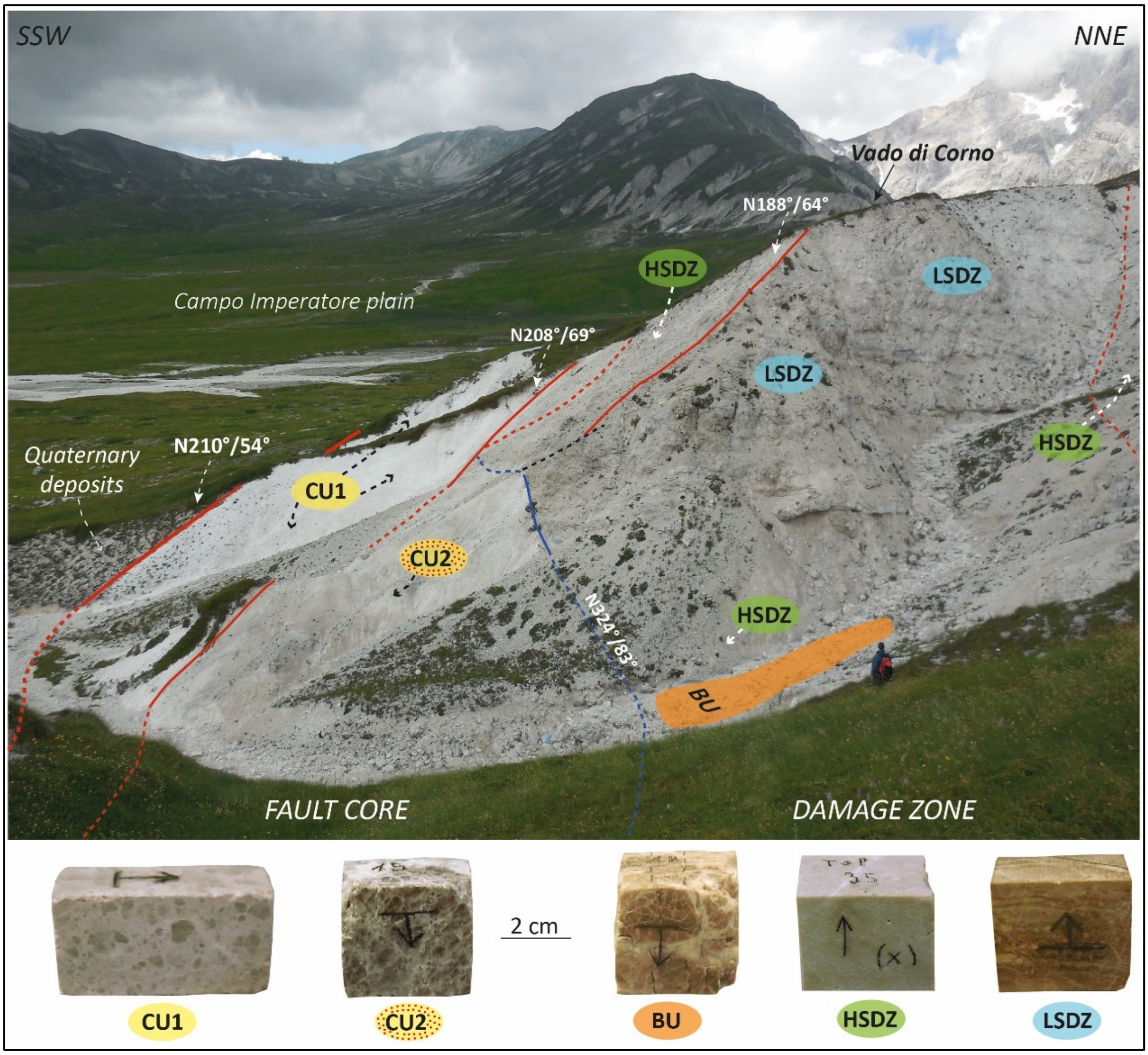

**Fig.3** (Upper side) Field view of the VCFZ footwall block, with the indication of the different structural units (CU1, CU2, BU, HSDZ, LSDZ), principal fault and high-hierarchy subsidiary faults (dashed parts are inferred) with attitude (N azimuth dip direction/ dip angle); see Fig.2 for the view point of Fig.3. Compare this view with the map and the structural section of Fig.2. See also Fig.S1 for a larger panoramic view. (Bottom side) Oriented rock samples representative of each distinct structural unit used for ultrasonic and petrophysical measurements. See photographs of all samples in Fondriest et al. (2024).

### 3.2. Ultrasonic velocity measurements

Ultrasonic P- and S-wave velocities were determined in transmission using a benchtop setup at the Rock Physics laboratory of the University College of London. The benchtop setup was equipped with a JSR DPR300 ultrasonic pulser/receiver with 35 MHz bandwidth, an Agilent Technologies 'Infiniium' digital oscilloscope with 1.5 GHz bandwidth and up to 8 GHz sampling rate, one pair of Panametrics V103 P-wave transducers and one pair of Panametrics V153 S-wave transducers with 1 MHz resonant frequency and 13 mm diameter piezoelectric elements (see Text S1 for more details). The signal sent to the transmitting transducer was a negative spike with amplitude of - 475 V, duration of 10-70 ns and pulsing frequency of 200-600 Hz. Before being sent to the oscilloscope, the signal coming from the receiving transducer was amplified and low-pass filtered to 3 MHz to improve the signal to noise ratio. Receiver waveforms were digitally recorded at the oscilloscope as stacks of 1024 individual times series with a sampling rate of 4 GHz. Transducers and specimen were mounted in a clamp assembly applying a vertical load of 0.1 kN (axial stress of ~ 0.8 MPa). As coupling agents, a thin layer of fluid ultrasonic gel was used for P-wave measurements and a high viscosity one for S-wave measurements. The first breaks of ultrasonic waves have been manually picked on screen using the software of the oscilloscope and picking uncertainties were ≤0.12 µs for P-waves and ≤0.25 µs for S-waves. Travel times of P- and S-waves through the system (i.e. rock sample plus transducer wear plates) were corrected for transducer delays (see Text S1). Four travel times were determined along each measuring direction by rotating the sample of 90° around the pulsing axis. The uncertainty associated to each velocity measurement was determined as the larger value between the error propagated from the travel path and the travel time uncertainties and the standard deviation of the four measurements along the same direction. This procedure guaranteed to catch any velocity variations due to sample internal heterogeneities, which could be particularly relevant for the polarized S-waves. The final velocity errors were ≤6% for P waves and ≤10% for S waves (Table S1). P-wave velocity anisotropy was calculated as difference between the highest and lowest directional velocity normalized by their average value and the associated uncertainty was derived by errors propagation.

**Fig.4** Field pictures of the VCFZ structural units. (a) Contact between the high-strain cataclastic unit CU1 (left side) and the lower strain cataclastic unit CU2 (right side) by an extensional subsidiary fault (red dashed line, fault plane dipping to the SSW). The CU1 accommodates larger bulk shear deformation compared to the CU2 (e.g. the dark grey lenses represent volumes of host dolostones with little shear deformation). (b) The breccia unit (BU), is cut by a mesh of dolomite veins. (c) High strain damage zone (HSDZ) affected by subsidiary faults and fractures spaced few centimeters apart. (d) Zoom of the HSDZ fracture network. (e) Lens of low strain damage zone (LSDZ) with recognizable bedding, larger fracture spacing and subsidiary faults spaced few meters apart. (f) Zoom of LSDZ outcrop, note the sub-horizontal bedding and the high-angle fractures spaced few tens of centimeters (see section 2.).

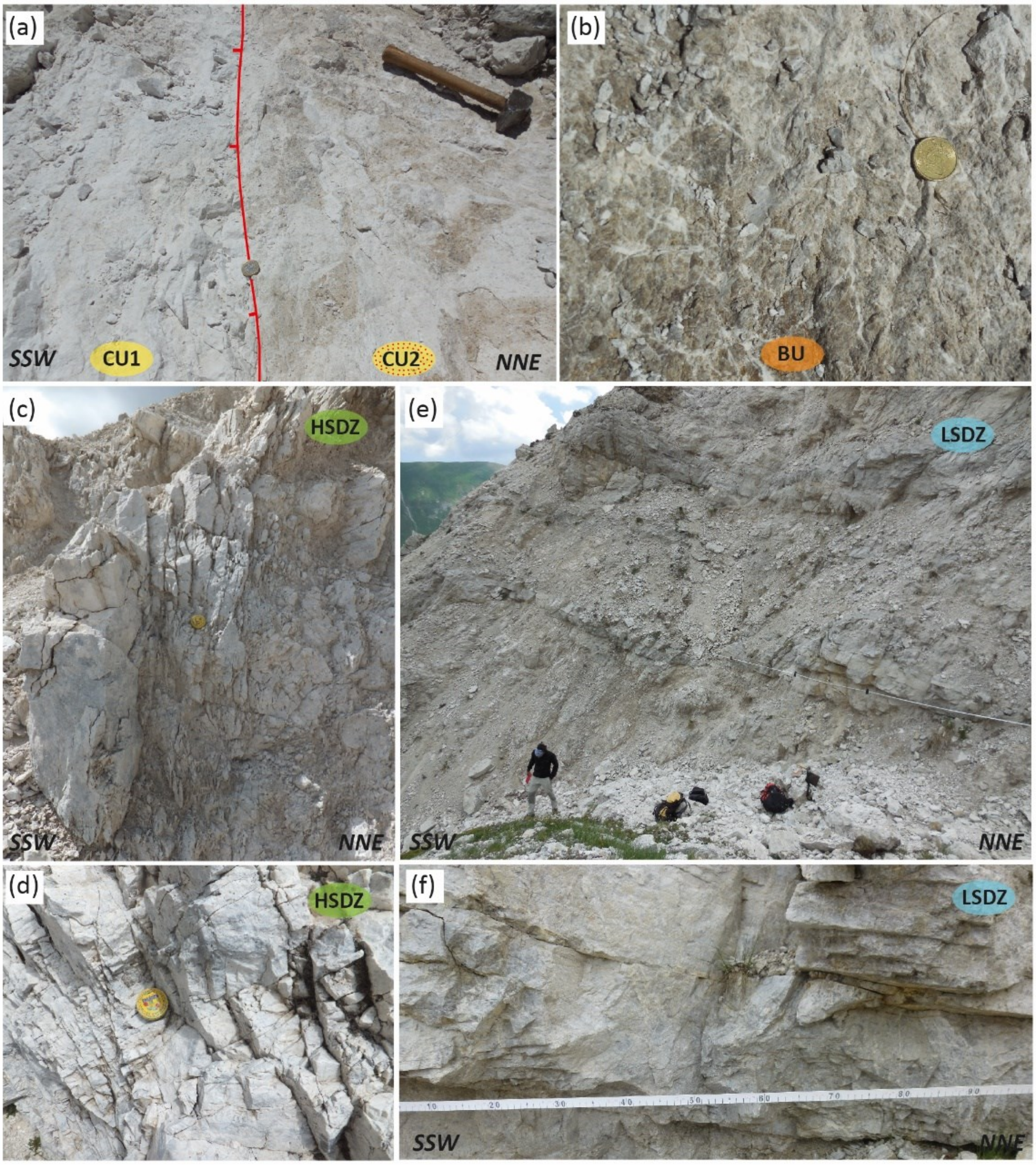

### 3.3. Density and porosity measurements

The solid (or grain) volumes of the oriented fault zone rock samples (n=30) were measured with two gas-displacement helium (He) pycnometers at the University College of London (AccuPyc 1340 II micromeritics) and at the University of Parma (Quantachrome Ultrapyc 1200; see Text S1). Before each measurement the instrument automatically purged water and volatiles from the sample. Multiple measuring cycles (N≥6) were performed until they converge to a consistent value; errors were determined as standard deviation of the different measurements. The solid (or grain) density of the samples was calculated as the ratio between the samples mass and solid volumes. Sample masses were measured with an analytical scale with a precision of $\pm10^{-4}$ g and errors on solid densities were propagated from those of masses and solid volumes. Bulk volumes and densities of the same fault zone samples were constrained using a modified wax-clod method (Blake and Hartge, 1986; Hao et al., 2008). After solid volumes have been determined, the samples were covered with an impermeable layer of paraffin wax. The sealed samples were weighed again, and the volume of the wax coat was estimated from the weight difference and the density of the wax of ~900 kg/m$^3$ at 20°C. The volume of the sealed samples was measured by He-pycnometry and the bulk volume was simply derived by subtracting the volume of the wax. Eventually the total connected porosity of the samples was calculated as percentage of the difference between the sample bulk and solid volume. Errors on porosities were ±0.01-0.38% and were propagated from those of bulk and solid volumes (Table 2).
Complementary porosity measurements were performed on thirteen fault zone samples by Hg-intrusion porosimetry. The measurements were performed at the University of Parma with a PoreMaster 33 system (Quantchrome Instruments) which allowed us to determine both total porosity and pore size distributions. In mercury porosimetry analysis, the volume of mercury penetrating into porous samples can be measured as a function of the applied hydraulic pressure. The obtained mercury intrusion and extrusion curves were interpreted into pore size distributions using the Washburn equation (Washburn, 1921; see Text S1 for more details). Before measurement, each sample was dried at 40°C for 24 h, and then ~1.5-2 g of sub-sample were analyzed in a sample cell 1.0 × 3.0 cm. Parameters used for measurements were: pressure range 0-228 MPa, mercury contact angle of 140°, and surface tension of mercury 0.48 N/m. Analyzed pore sizes were in the range 0.0064 to 500 μm and total porosity errors were ±0.01-0.02% (see Text S1 for more details).

### 3.4. Microstructural analyses

A total of fifty-seven polished thin sections (33x20 mm, 0.03 mm thick) were cut from the fault zone rock samples for which ultrasonic velocities and He-pycnometer porosity were measured. In particular, three mutually orthogonal thin sections were obtained from each fault-oriented sample and few others from the bases of rock slabs and cylinders orthogonal to the vertical direction. The thin sections were stained with Red Alizarin-S to easily distinguish calcite and dolomite. The thin sections were studied on high-resolution scans and at the optical microscope in transmitted light at the Institut des Sciences de la Terre (ISTerre) in Grenoble. Few selected thin sections were investigated in scanning electron microscopy (SEM) at the Institut des Sciences de la Terre (ISTerre) in Grenoble and at the Department of Geosciences of the University of Padova, using respectively a Vega3 Tescan (voltage 10 kV) and a field emission SEM Tescan SOLARIS (10 kV) in backscattered and ultra-high-resolution imaging mode. These analyses allowed us to characterize the mineralogy and the microstructures (e.g. cataclastic versus fracture-dominated fabrics, porosity and cement distribution) of the fault zone rocks in different orientations and to look for directional changes relating to the ultrasonic velocity measurements.

### 3.5. High-resolution P-wave seismic survey and tomography inversion

One active-source P-wave seismic survey was performed along a ~90 m long profile across the VCFZ (profile t-t' in Figs.2-S2). The acquisition line was equipped with 72 vertical geophones (natural frequency at 4.5 Hz) equally spaced by ~1 m, which were handled by three multi-channel seismic stations arranged in series (24 channels Geometrics Geode). To generate the seismic signals, a 5 kg sledge hammer was hit vertically on a metal plate connected to an integrated trigger system. The source was moved in between the geophones along the line with a shot spacing of 3 m. Additional shots were performed at -10 m and -5 m from the first geophone, and at +5 m and +10 m from the last geophone. The geophone line was connected to a field computer, which allowed viewing of seismic recordings related to each shot. Each acquired trace had a total duration of 1.5 seconds and contained digitized seismic data with a sampling step of 0.125 ms (8000 Hz). For first-arrival time readings, the seismic traces were organized into 28 common shot seismic sections obtained by arranging the seismograms in sequence according to source-receiver distance (Fig.S3). First breaks picking of P-waves were performed manually on the seismic sections with the traces normalized for their maximum amplitude value. Out of the 2016 available traces, 1791 P-wave first arrival times were manually measured avoiding to pick the noisiest traces. The picking uncertainties were on the order of 2 ms (Figs.S3-4). Due to the irregular topography of the area, the spatial coordinates of both the shots and the receivers were measured with a GPS system based on the RTK (Real-Time Kinematic) technique with an accuracy of few centimeters. The precise topography reconstruction was essential for the tomographic inversion.

First arrival times of P-waves were inverted using a non-linear tomographic code coupled with a multiscale inversion strategy (for details see Herrero et al. 2000; Improta et al. 2002). The inversion of first arrivals was performed on the basis of a comparison between the data calculated on a given model and those measured on the seismic recordings. For a fixed propagation medium, parameterized through a regular grid of nodes defining a bi-spline, the data were computed by solving the Eikonal equation with a finite difference method (Podvine and Lecomte, 1991). The problem of determining the velocity model was addressed by using a mixed global/local optimization method based on Monte Carlo and Simplex methods, respectively. For the determination of the final P-wave velocity model we used a multiscale (or multistep) approach based on the gradual increase of the number of nodes starting, at each step, from the model defined by a smaller nodes number obtained in the previous step. This type of strategy makes robust the model parameters estimation and stabilizes the convergence of the inversion process. Starting from a very simple model defined by only 2 nodes, we gradually increased the number of nodes arriving at the construction of the final seismic model parameterized by 84 nodes (12 along horizontal axis and 7 along the vertical axis). The root mean square (RMS) difference between the measured times and those theoretically calculated on the final model was 2.4 ms, comparable to the measurement error on the data (Fig.S5). The path and the density of the seismic rays passing through the investigated model was used to determine the areas of the final model that were well sampled (i.e. spatially well resolved), and thus well constrained during the inversion. The distribution of ray paths obtained both in the final model and intermediate models (defined by a smaller number of nodes) was used to determine this area. For the intermediate models, the investigation depth of rays was halved in order to differently weigh the contribution made by the different models in defining the resolved areas, thereby giving greater weight to the rays' distribution defined in the final model. The seismic survey was conducted during the summer period in a dry karstic area with exposed intensely fractured carbonates. Therefore, there was no effect of water saturation on the derived seismic velocities.

### 3.6. Outcrop scale measurement of fractures

Fracture abundance in the VCFZ was determined at the mesoscale on outcrops of the fault core units CU1-CU2 and of the extensional damage zone units HSDZ and LSDZ (see Fig.5). It was computed as areal fracture intensity ($m^{-1}$), defined as length of fracture traces per unit scan (or sampling) area (P21 *sensu* Dershowitz and Herda, 1992). Since P21 depends both on the size of the scan area and its orientation, fractures were measured on scan areas with comparable sizes and oriented perpendicular to the principal fault surface (only for the HSDZ two fault parallel scan areas were also realized). The scan areas range from 60x40 $cm^2$ to 80x100 $cm^2$. Fracture tracing was performed on outcrop photographs (pixel size ≤1 mm) acquired with a digital camera and then orthorectified (see text S2 for more details). Fracture traces were digitized using the software ArcMap 10.7.1 which allowed us to precisely calculate their length together with the size of the scan area corrected for any covered portion. To prevent biases due to the resolution of the techniques (i.e. data truncation) and the finite size of the scan rea (i.e. data censoring), P21 was calculated on a selected range of fracture length. Such range correspond to the length interval where the fracture size distribution of a specific outcrop was displaying a linear trend in a logarithmic plot. The 2D fracture distributions thus determined were used to calculate, through an effective medium approach, the reduction of the elastic moduli in the VCFZ structural units from the ultrasonic length scale (2-5 cm) to the scale of the outcrop (1-2 m; see section 5.2 for details). With the aim to consider the effect of rock mass discontinuities on elastic properties in the first 10-20 m depth (0.25-0.5 MPa vertical load), we count as fractures: faults (dominant in CU1-CU2), shear fractures with negligible slip, sets of joints (dominant in HSDZ and LSDZ), and bedding surfaces in the LSDZ as continuous interfaces affected by pressure-solution. In contrast, the sealed microfractures dominating CU1-CU2 and the veins in the damage zone units were not considered.

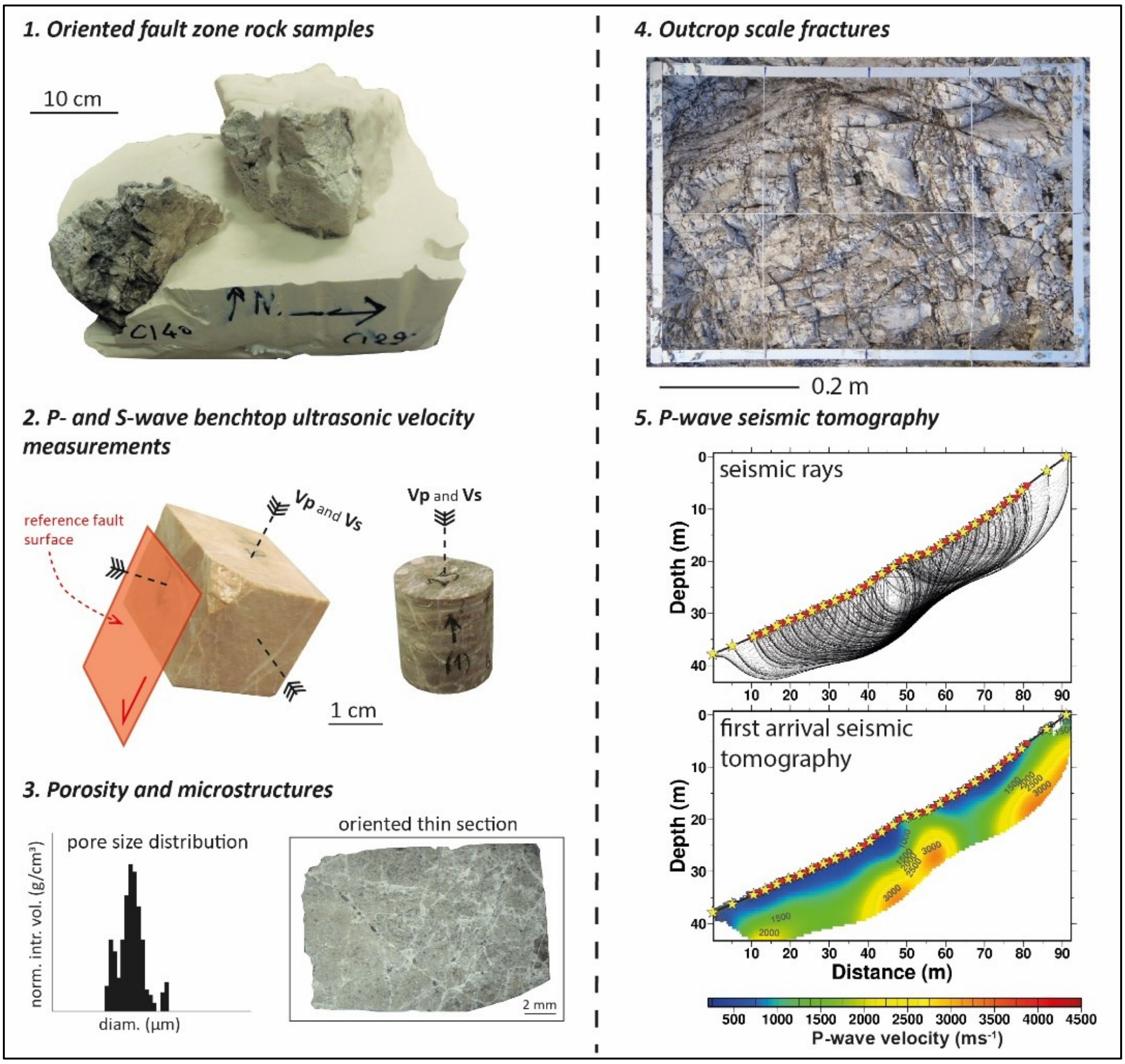

**Fig.5** Laboratory- and field-scale methods. 1. Orientation of fault zone rock blocks; 2. Production of few centimeters in size samples oriented with respect to faults (mutual orthogonal faces) or to the topographic vertical. Directional P- and S- ultrasonic velocity were measured under axial load of ~0.1 kN. 3. Bulk porosity (He-pycnometry), pore size distribution (Hg-porosimetry) analyses and microstructural analyses on oriented mutually orthogonal thin sections. 4. Outcrop scale fracture intensity measurements (P21 sensu Dershowitz and Herda, 1992). 5. Field-scale P-wave first arrival seismic tomography (seismic Vp in $ms^{-1}$). Seismic ray density and seismic velocity distribution with depth under the seismic profile. See section 3. of the main text for more details.

## 4. Results

### 4.1. Ultrasonic wave velocities across the fault zone

Ultrasonic P- and S elastic waves were measured on the oriented fault zone rock specimens at increasing distance from the principal fault surface (Fig.6). The different structural units (CU1, CU2, BU, HSDZ, LSDZ) have been regularly sampled along a ~200 m long profile with the aim to characterize any significant spatial variation of elastic properties (Fig.2). In Fig.6a, the P- and S-wave velocities of each sample are represented as average of the three mutual orthogonal measurements (see section 3.1.), while each single directional measurement, including those along the vertical, is reported in Table S1. As the distance from the main fault surface increases, no gradual change in average ultrasonic wave velocities has been observed. Instead the fault zone is characterised by the juxtaposition of structural units marked by different ranges of ultrasonic velocities (Fig.6a). In particular, the fault core cataclastic units CU1-CU2 (map distance from principal fault surface: 0-32 m) show average P-wave velocities of 4-5 $kms^{-1}$ and S-wave velocities of 2.2-3 $kms^{-1}$. These values are significantly lower compared to the ultrasonic velocities of the high- and low-strain damage zone units (HSDZ-LSDZ; map distance from principal fault surface: 50-160 m), with average P-wave velocities of 5-6.5 $kms^{-1}$ and S-wave velocities of 3-3.6 $kms^{-1}$ (Fig.6a). The breccia unit (BU), which was sampled at the few discontinuous exposures within the extensional damage zone, from its internal border up to ~ 100 m distance, shows intermediate velocity values (Fig.6a). Importantly, the fault core units are affected by a strong dispersion of velocity values, both for P- and S-waves (Fig.6a-b), and the HSDZ comprises few samples (e.g. *samples 29 and 32* in Fig.6a) with very reduced velocity (P-wave ~ 4.5 $kms^{-1}$ and S-wave ~ 2.7 $kms^{-1}$). These two relevant points will be addressed below in relation to samples variability in terms of porosity and microstructures (sections 4.1.-4.3.).

Focusing on the proximal fault core unit CU1, which accommodates most of the fault zone shear deformation, Fig.6b displays the directional ultrasonic P- and S-wave velocities of each sample at increasing distance from the principal fault surface (0-25 m). Both P- and S-waves do not show any correlation with distance and mean velocity values range from ~3.7 to 5.1 $kms^{-1}$ for P-waves and 2.2 to 3 $kms^{-1}$ for S-waves (Fig.6b). Velocity variations among single fault core samples are generally larger than the internal directional velocity variations, suggesting a significant sample variability possibly associated to textural heterogeneities within the fault core. Ultrasonic P-wave velocity anisotropy reaches mean values ≤12% in CU1 with 44% of samples slower in the fault perpendicular direction, 44% in the fault parallel horizontal direction and 12% in the fault parallel vertical direction (Fig.6b, c). Measured anisotropy on S-waves are smaller and no significant variations were detected by measurements with wave polarity rotations of 90° (i.e. see the small error bars in S-wave data of Fig.6b). P-wave anisotropy data across the entire fault zone are represented in Fig.6c with a maximum mean value of ~18% for a CU2 sample. The slowest direction is the fault perpendicular one for 45% of the samples, the fault parallel horizontal for 30% of the samples and the fault parallel vertical for the remaining 25%. In the LSDZ, specimens from the fault-bounded lens within the HSDZ (see light blue unit in Figs.2, 3, 4e, f), show anisotropy up to ~10% and the slowest direction is the fault parallel vertical one at high angle to the bed-parallel laminations affecting the rocks at the sub-centimetre scale (see LSDZ sample in Fig.3). For LSDZ samples in the external part of the fault zone (samples CIG40-41 in Fig.1 and grey data in Fig.6c), the sensitivity of ultrasonic velocity has been investigated with respect to the orientation of two mesoscale joint sets affecting the rock mass. The two joint sets strike N-S (J1: N84/76) and E-W (J2: N358/81) with spacing of few tens of centimetres. The measured P-wave anisotropy ranges from ~2% to ~13% with the slowest direction orthogonal to J2 for both samples (see Table S1).

!

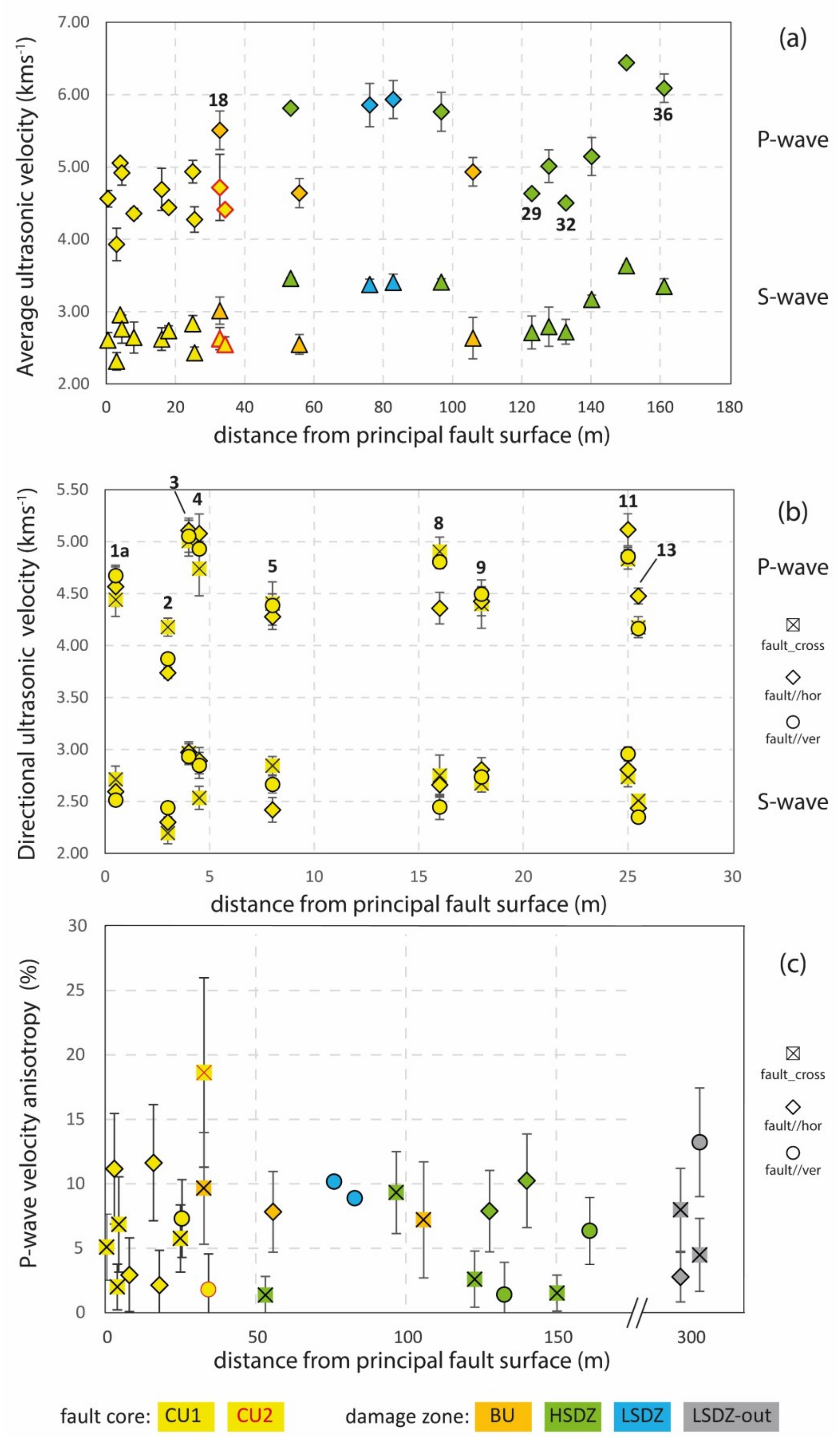

**Fig.6** (a) Average ultrasonic P- and S-wave velocities (diamonds and triangles respectively) of the oriented rock samples within the VCFZ at increasing distance from the principal fault surface. The color code is reported at the base of the figure. The number (*n*) of few samples (sample name is *CIG n*) described in the main text and shown in Fig.10 are reported. (b) Directional ultrasonic P- and S-wave velocities for rock samples of the high-strain cataclastic unit CU1 at increasing distance from the principal fault surface. Number (*n*) of CU1 samples (sample name is *CIG n*) described in the main text and shown in Fig.9 are reported. (c) P-wave velocity anisotropy of the oriented rock samples across the VCFZ at increasing distance from the principal fault surface. The symbols indicate the slowest among the measured three mutually orthogonal directions. See section 4.1. of the main text for a detailed data description.

**4.2. Ultrasonic wave velocities versus porosity**

The relations between ultrasonic P- and S-wave velocities and total connected porosity (He pycnometer data) were investigated for all the oriented samples (Figs. 7, S6-S8). In Fig.7, ultrasonic P- and S-velocities along the fault perpendicular direction (i.e. the slowest for most of the samples) are displayed, while velocities in the other two mutual orthogonal directions and along the vertical are shown in Figs. S6-8 respectively. Samples from the proximal fault core (i.e. high shear deformation – unit CU1) display connected porosities ranging from ~3% to ~10%, while samples from all the other structural units have porosities ≤2-2.5% (Table 2; Figs. 7, S6-8). A general negative correlation exists between ultrasonic velocities and total porosity. More specifically, the trends of directional velocities are fairly well described by exponential decays with increasing porosities, similarly for P- and S-waves (Fig. 7, S6-8). Clear outliers from these trends, characterized by very low velocities, are represented by *samples 29 and 32* (HSDZ), *42* (BU), *and 16* (CU2) (see data in Figs. 7, S6-7). In addition, individual samples from the proximal fault core (CU1) show uncorrelated ultrasonic velocities: e.g. *samples 2* has both lower velocities and porosities compared to *sample 9* and *4* (Figs.7, S6-7), as well as *sample 8-ver* compared to *sample 5* (Fig.S8).

Complementary measurements by Hg intrusion porosimetry (section 3.3.) yielded total porosities comparable with He-pycnometry (Fig.8). The measurements overlap particularly well for the samples with total connected porosity >2% (Table 2). The structural units are marked by different pore size distributions (Fig.8). In particular, the more porous proximal fault core (high shear deformation – unit CU1) displays a narrow peak maximum at ~0.2 µm and relative maxima at ~0.06 and ~200 µm (median pore sizes of 0.19-0.23 µm; Fig.8). The breccia unit (BU) shows a lower magnitude and broader maximum centred at ~0.3 µm, while the high-strain damage zone (HSDZ) is characterized by the same broad relative maximum and peak maxima at ~8 µm and 200-300 µm (Fig.8). The low-strain damage zone (LSDZ) displays pore size distributions with no clear relative maxima (Fig.8).

| Sample | Struct. unit | Porosity He (%) | st. dev. | Porosity Hg (%) | st. dev. | Bulk vol. ($cm^3$) | st. dev. | Bulk den. ($g/cm^3$) | st. dev. |
|---|---|---|---|---|---|---|---|---|---|
| CIG1a | CU1 | 6.65 | 0.09 | | | 4.95 | 0.002 | 2.64 | 0.001 |
| CIG2/CIG2_ver | CU1 | 4.85 | 0.29 | 4.94 | 0.02 | 26.12 | 0.058 | 2.63 | 0.006 |
| CIG3-CIG3_ver | CU1 | 2.23 | 0.38 | | | 20.08 | 0.039 | 2.62 | 0.005 |
| CIG4 | CU1 | 7.51 | 0.04 | | | 3.73 | 0.001 | 2.58 | 0.001 |
| CIG5/CIG5_ver | CU1 | 4.59 | 0.25 | 4.64 | 0.01 | 8.38 | 0.001 | 2.66 | 0.000 |
| CIG8 | CU1 | 2.70 | 0.06 | 3.21 | 0.01 | 12.87 | 0.006 | 2.71 | 0.001 |
| CIG8_ver | CU1 | 3.04 | 0.08 | | | 14.45 | 0.007 | 2.71 | 0.001 |
| CIG9 | CU1 | 5.34 | 0.06 | | | 15.42 | 0.004 | 2.63 | 0.001 |
| CIG11 | CU1 | 3.86 | 0.15 | | | 21.58 | 0.022 | 2.69 | 0.003 |
| CIG13 | CU1 | 9.71 | 0.36 | | | 38.14 | 0.142 | 2.67 | 0.010 |
| CIG14 | BU | 2.28 | 0.03 | 1.09 | 0.02 | 4.42 | 0.000 | 2.73 | 0.000 |
| CIG15 | CU2 | 1.72 | 0.13 | | | 10.53 | 0.014 | 2.76 | 0.004 |
| CIG16 | CU2 | 1.43 | 0.18 | | | 12.79 | 0.024 | 2.76 | 0.005 |
| CIG17 | BU | | | 0.99 | 0.01 | | | | |
| CIG18 | BU | 1.3 | 0.02 | 0.51 | 0.01 | 21.83 | 0.002 | 2.79 | 0.000 |
| CIG18_ver | BU | 0.14 | 0.02 | | | 9.27 | 0.001 | 2.79 | 0.000 |
| CIG19 | HSDZ | 0.65 | 0.03 | | | 30.66 | 0.008 | 2.81 | 0.001 |
| CIG24 | LSDZ | 1.78 | 0.01 | 0.51 | 0.02 | 27.77 | 0.002 | 2.78 | 0.000 |
| CIG24_ver | LSDZ | 1.20 | 0.12 | | | 14.27 | 0.008 | 2.80 | 0.002 |
| CIG25 | LSDZ | 1.76 | 0.04 | 0.93 | 0.02 | 16.61 | 0.005 | 2.77 | 0.001 |
| CIG25_ver | LSDZ | 0.82 | 0.05 | | | 6.86 | 0.001 | 2.80 | 0.001 |
| CIG26 | LSDZ | | | 0.97 | 0.01 | | | | |
| CIG27 | HDSZ | 1.24 | 0.09 | 0.84 | 0.02 | 19.13 | 0.010 | 2.80 | 0.001 |
| CIG27_ver | HDSZ | 0.84 | 0.02 | | | 15.05 | 0.002 | 2.80 | 0.000 |
| CIG29 | HDSZ | 1.11 | 0.02 | 1.08 | 0.01 | 45.84 | 0.011 | 2.78 | 0.001 |
| CIG29_ver | HDSZ | 0.12 | 0.02 | | | 14.68 | 0.002 | 2.81 | 0.000 |
| CIG30 | HDSZ | | | | | | | | |
| CIG32 | HSDZ | 1.60 | 0.05 | | | 17.99 | 0.008 | 2.78 | 0.001 |
| CIG33 | HDSZ | 1.27 | 0.03 | | | 39.99 | 0.011 | 2.77 | 0.001 |
| CIG34 | HDSZ | | | 1.25 | 0.01 | | | | |
| CIG35 | HDSZ | 0.35 | 0.04 | | | 25.56 | 0.008 | 2.80 | 0.001 |
| CIG36 | HDSZ | 0.49 | 0.03 | | | 13.92 | 0.001 | 2.79 | 0.000 |
| CIG40 | LSDZ | 0.76 | 0.03 | | | 13.46 | 0.003 | 2.80 | 0.001 |
| CIG42 | BU | 0.88 | 0.03 | 0.65 | 0.00 | 27.06 | 0.005 | 2.78 | 0.001 |

**Table 2.** Data of He-porosity, Hg-porosity, bulk volume and bulk density of the analysed fault zone rock samples.

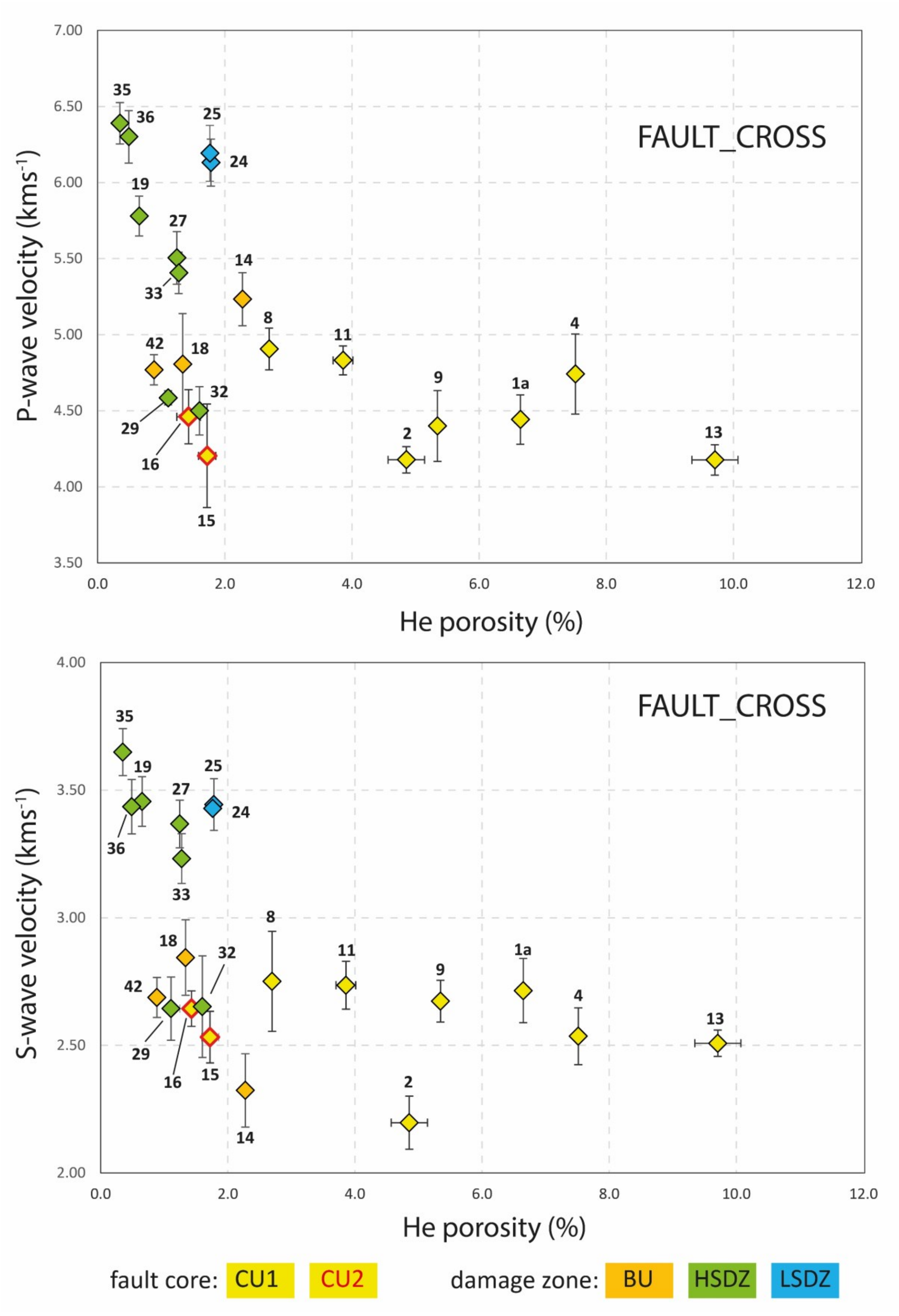

**Fig.7** Ultrasonic P- and S-wave velocities in the fault perpendicular direction (fault_cross) versus He-porosity data. P-wave velocities are reported in the upper plot and S-waves in the lower one. Number (*n*) of samples (sample name is *CIG n*) are reported. See section 4.2. of the main text for a detailed data description.

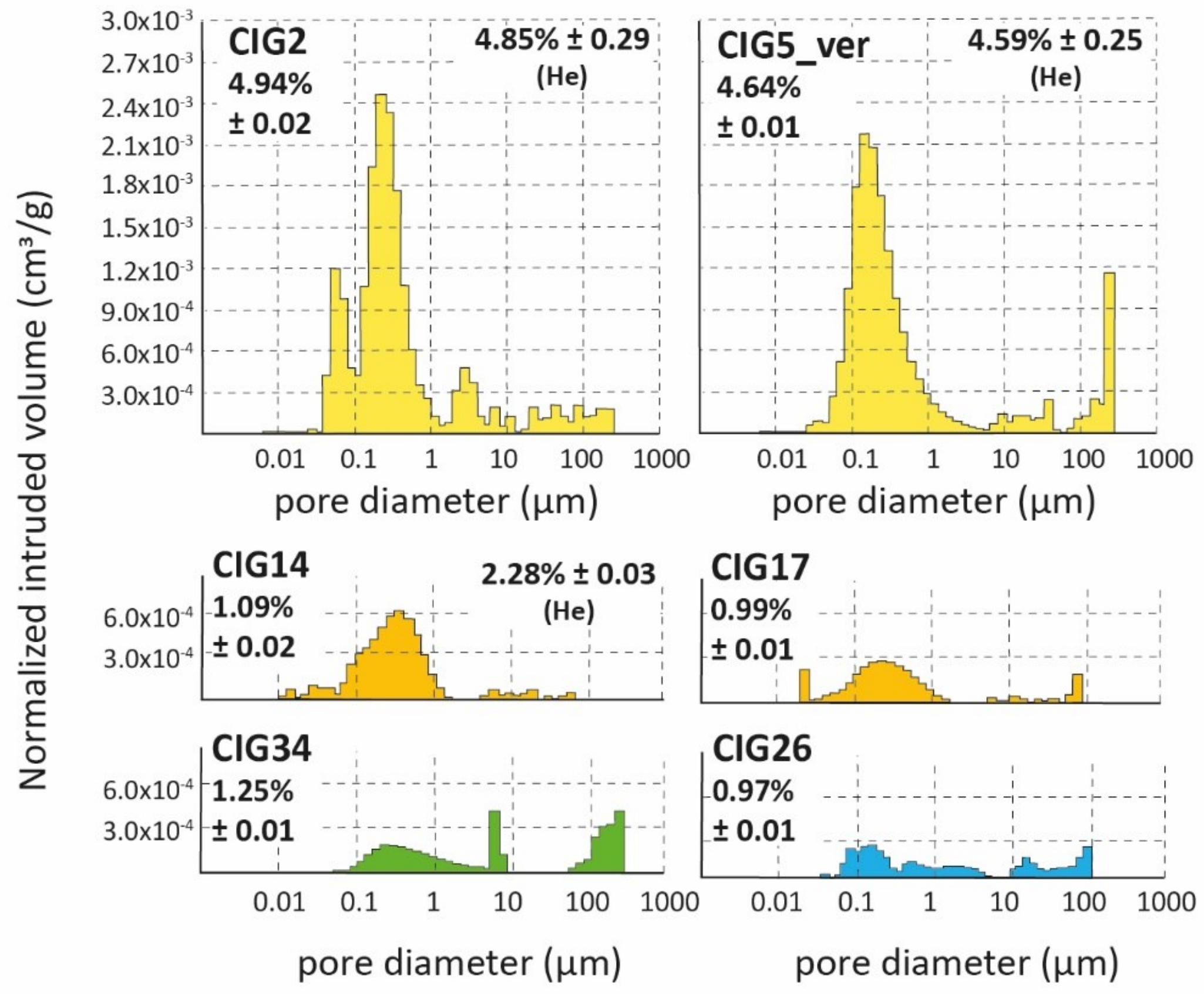

**Fig.8** Hg-porosimetry data. Fault core CU1 (*CIG2* and *CIG5_ver*; color: yellow), breccia unit (*CIG14* and *CIG17*; color: orange), HSDZ (*CIG34*; color: green) and LSDZ (*CIG26*; color: light blue) fault zone rock samples. Pore size distributions are reported as histograms (normalized intruded Hg volume on the y-axis versus pore diameter on the x-axis), while total porosity from Hg-porosimetry and He-pycnometry are reported in upper left and upper right corners of each plot respectively. See section 4.2. of the main text for a detailed data description.

### 4.3. Physical properties and microstructures of the fault zone rocks

The analysis of the oriented fault zone rocks at the thin section scale allowed us to relate elastic properties with mineralogy and microstructures in mutual orthogonal directions (Figs.9-10). In Figs.9-10 specific oriented thin section scans were selected to disclose the variability of elastic properties within the fault core and the characteristics of the outliers (*samples 29 and 32*) in the high-strain damage zone (HSDZ). All the analysed thin section scans are shown in Fondriest et al. (2024).

At the investigated length scale, the proximal fault core (CU1) is characterized by domains with different textural maturity regardless of the distance from the principal fault surface (Fig.9). The fault rock microstructures span from protocataclasites (e.g. Fig.9a-b), to cataclasites (Fig.9c) and ultracataclasites (Fig.9d). The fault rock mineralogy, determined both in optical microscopy and SEM, includes dolomite from the host rock and calcite which can locally reach up to ~40% in ultracataclasites (2D estimates based on SEM images; Fig.9d). Protocataclasites are affected by limited shear strain and display a clast-supported texture with fine-grained dolomite matrix localized along main fractures (Figs.9a-b). Connected porosity is both intergranular and along fractures (intra- and transgranular ones). Fractures are partially filled by calcite cements and small calcite veins are locally present (calcite <10-15% in 2D; Fig.9a-b). P-wave anisotropy increases with the occurrence of large continuous fractures and calcite veins (*sample 2 versus sample 5* in Fig.9). Cataclasites are characterized by a more isotropic fabric (rounded dolomite clasts surrounded by abundant matrix) and pronounced sealing of the intergranular pore space by calcite cement resulting in lower porosities, P-wave anisotropies and higher P-wave velocities (Fig.9c). Ultracataclasites are characterized by the highest percentage of matrix and calcite in the intergranular pore space (up to 40% in 2D; Fig.9d). The resulting P-wave velocities are comparable with those of protocataclasites as well as to cataclasites, even though connected porosities are much larger: ~10% for *sample 13* compared to ~4% for *samples 2 and 5*, and ~7.5% for *sample 4* compared to ~4% for *samples 11* (Figs.7, S6-7).

Fault zone rocks from the extensional damage zone includes samples of HSDZ, BU and LSDZ (Fig.10). At the microstructural level, samples from the high-strain damage zone (HSDZ) consist of strongly dolomitized platform carbonates affected by millimetres thick dolomite veins formed during the older thrust-related deformation phase (Figs.10a-b). P-wave velocities reach upper bounds > 6 $kms^{-1}$ and anisotropy is <10%. The HSDZ outliers (samples 29 and 32) with very reduced elastic wave speeds (P-wave ~ 4.5 $kms^{-1}$ and S-wave ~ 2.7 $kms^{-1}$), are instead affected by networks of both fault-parallel and perpendicular thin shear fractures formed during the later extensional deformation phase. These fractures display local cataclastic fabrics and are partially filled by calcite (Fig.10a). Connected porosity is intergranular along these fractures, while diagenetic vugs account for isolated porosity (Fig.10a). Samples from the breccia unit (BU) are characterized by intermediate elastic wave speeds and at the microscale they are affected by dense networks of thick dolomite veins which locally produce crush breccia textures with suspended clasts (Fig.10c). Depending on their structural position within the fault zone, some BU samples are also affected by thin calcite-filled shear fractures related to extension (Figs.6a). Samples of the low-strain damage zone (LSDZ) fault-bounded lens (light blue unit in Figs.2, 3, 4e, f) are dolomicrites with bed-parallel laminations and intraclasts rich layers. These samples are affected only by tiny shear fractures and pressure solution seams, and P-wave velocities are around 6 $kms^{-1}$.

**Fig.9** Physical properties and microstructures of the high-strain cataclastic unit CU1 (fault core). Each figure section contains a high-resolution scan of an oriented thin section stained with Red Alizarin-S and a SEM backscattered image of a zoomed area indicated in the thin section scan. The displayed thin sections are cut parallel to the principal fault surface and thus orthogonal to the fault perpendicular direction of ultrasonic velocity measurement. For each thin section are indicated the P- and S-wave velocities in the fault perpendicular direction, P-wave anisotropy and He-porosity measured on the bulk sample. (a-b) Thin sections *CIG2 fault //* and *CIG5 fault //* from samples *CIG2* (distance from principal fault surface d = 3 m) and *CIG5* (d = 8 m) respectively. Both samples show a protocataclastic microstructure with more comminuted clast-supported bands along main fractures. Connected porosity is intergranular along fractures and partially filled by calcite cement. Small calcite veins are locally present. (c) Thin section *CIG11 fault //* from samples *CIG11* (d = 25 m) shows a more isotropic cataclastic microstructure with rounded dolomite clasts surrounded by abundant matrix and pronounced sealing of the pore space by calcite cement. See the pervasive sealing in the SEM images and the remnant intergranular microporosity. (d) Thin section *CIG13 fault //* from samples *CIG13* (d = 26 m) shows an ultracataclastic matrix-supported microstructure associated with an increase of the intergranular porosity which is only partially sealed by later calcite cement. See section 4.3. of the main text for a detailed microstructural and petrophysical data description.

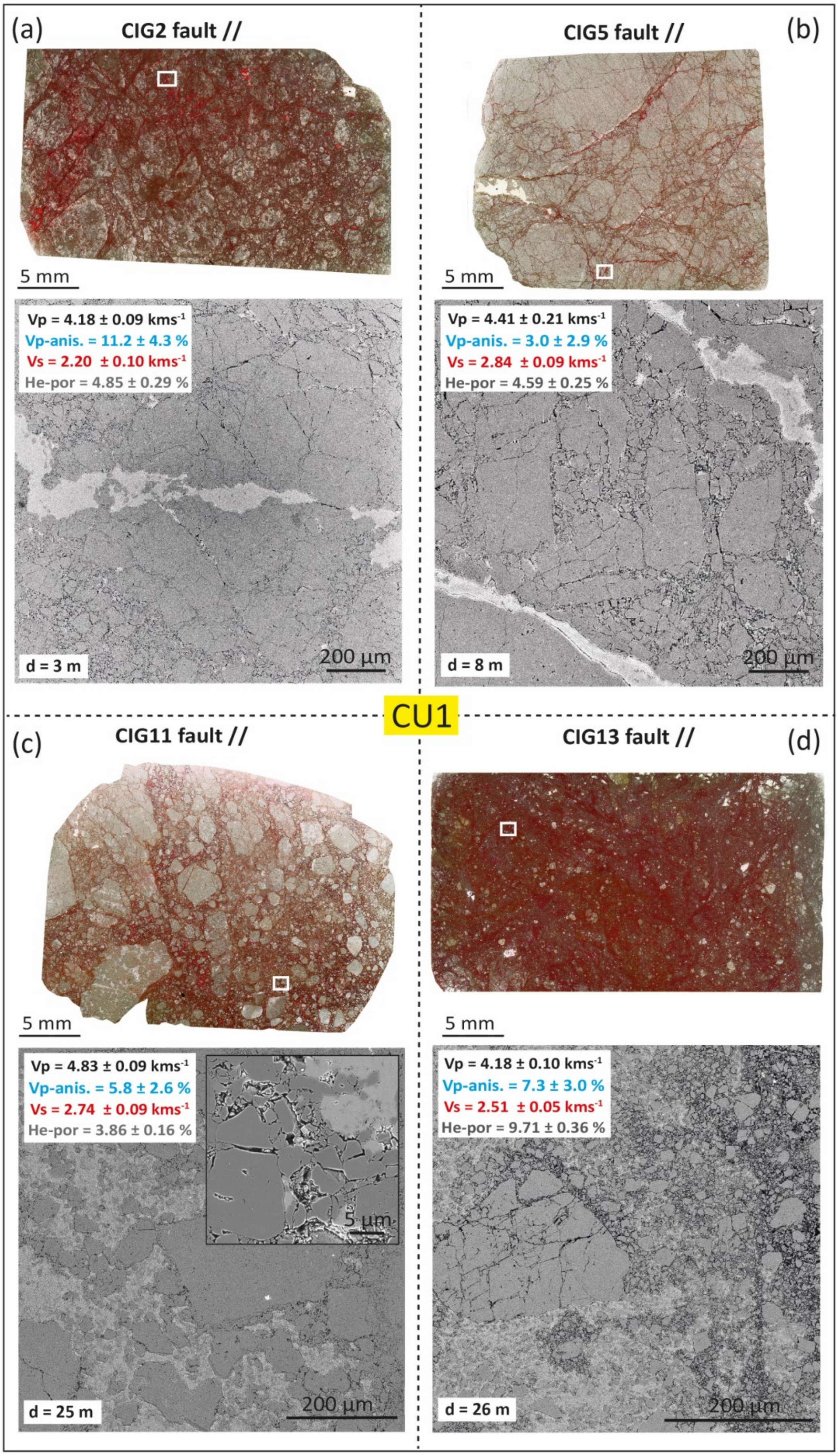

(a)
CIG2 fault //
5 mm
Vp = 4.18 ± 0.09 kms-1
Vp-anis. = 11.2 ± 4.3 %
Vs = 2.20 ± 0.10 kms-1
He-por = 4.85 ± 0.29 %
d = 3 m
200 µm
CIG5 fault //
(b)
5 mm
Vp = 4.41 ± 0.21 kms-1
Vp-anis. = 3.0 ± 2.9 %
Vs = 2.84 ± 0.09 kms-1
He-por = 4.59 ± 0.25 %
d = 8 m
200 µm
CU1
(c)
CIG11 fault //
5 mm
Vp = 4.83 ± 0.09 kms-1
Vp-anis. = 5.8 ± 2.6 %
Vs = 2.74 ± 0.09 kms-1
He-por = 3.86 ± 0.16 %
5 µm
d = 25 m
200 µm
CIG13 fault //
(d)
5 mm
Vp = 4.18 ± 0.10 kms-1
Vp-anis. = 7.3 ± 3.0 %
Vs = 2.51 ± 0.05 kms-1
He-por = 9.71 ± 0.36 %
d = 26 m
200 µm

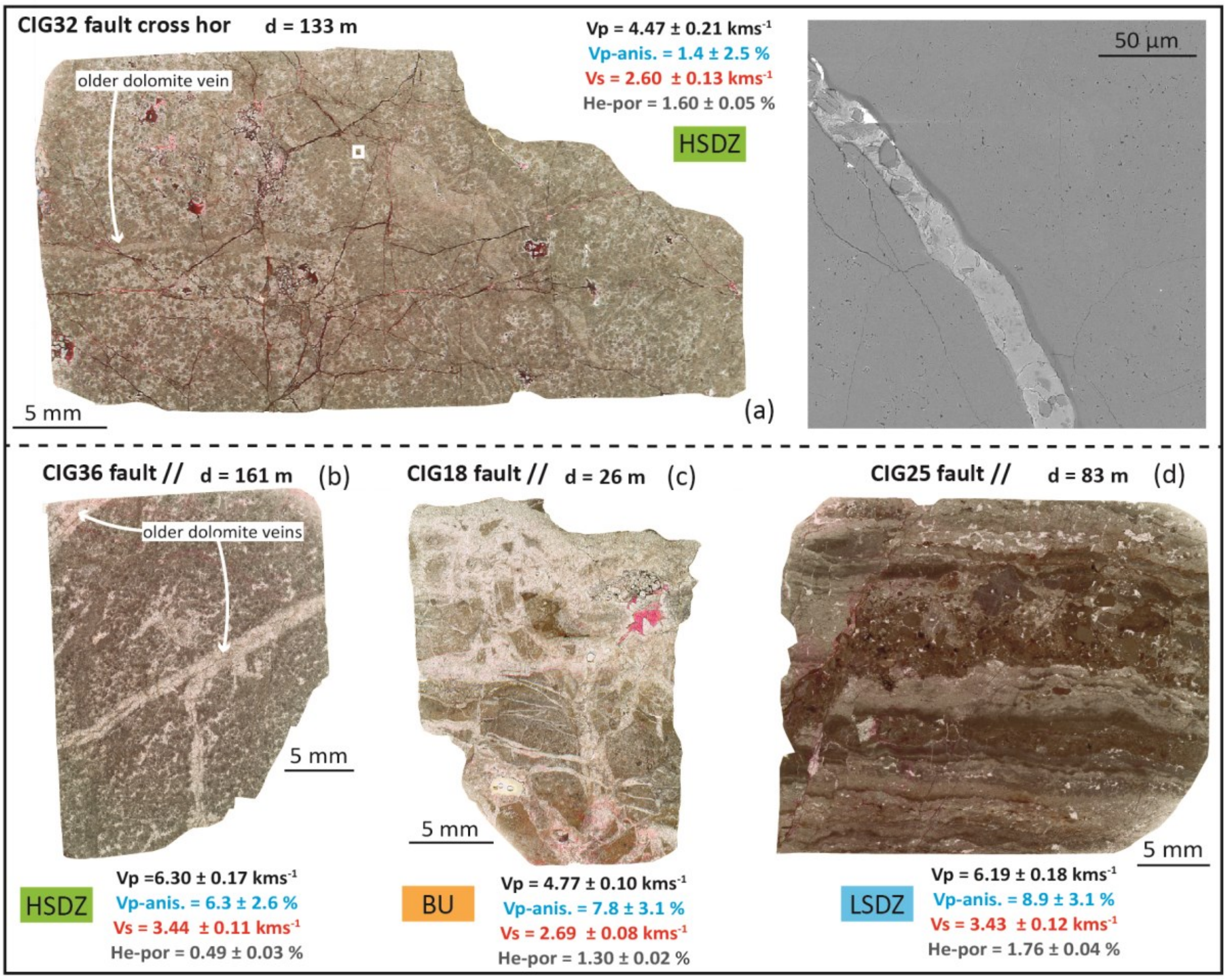

**Fig.10** Physical properties and microstructures of the VCFZ extensional damage zone including HSDZ, BU and LSDZ. Each figure section contains a high-resolution scan of an oriented thin section stained with Red Alizarin-S. The occurrence of calcite cement is significantly reduced compared with the fault core samples (Fig.9) and is only within extension-related microfractures. The displayed thin sections are cut parallel to principal fault surface and orthogonal to the fault perpendicular direction of ultrasonic velocity measurement, apart from panel (a). For each scan are indicated the P- and S-wave velocities in the direction perpendicular to the thin section, P-wave anisotropy and He-porosity measured on the bulk sample. (a) Thin section *CIG32 fault cross hor* from sample *CIG32* (d = 133 m) was cut orthogonal to the reference fault surface and parallel to its strike. It is representative of a HSDZ sample with very reduced ultrasonic velocities due to the presence of a network of fault-parallel and perpendicular shear fractures related to extension and partially filled by calcite cement (see the details of the calcite infill in the BSE-SEM image). The surrounding rock is a strongly dolomitized platform carbonate affected by older dolomite veins related to the older compressional deformation phase. (b) Thin section *CIG36 fault //* from sample *CIG36* (d = 161 m) is representative of the HSDZ with standard high ultrasonic velocity values. It is a strongly dolomitized platform carbonate cut by few millimeters thick dolomite veins formed during compression. (c) Thin section *CIG18 fault //* from sample *CIG18* (d = 26 m) is representative of the BU associated to the Omo Morto thrust zone. It is affected by a dense networks of thick dolomite veins which locally produce crush breccia textures with suspended clasts and shows intermediate ultrasonic velocities. (d) Thin section *CIG25 fault //* from sample *CIG25* (d = 83 m) is representative of the fault-bounded lens of LSDZ (light blue unit in Figs.2-4). It is a dolomicrite with bed-parallel laminations and intraclasts rich layers. Only tiny shear fractures and pressure solution seams are observed and ultrasonic velocities are high. See section 4.3. of the main text for a detailed microstructural and petrophysical data description.

### 4.4. P-wave seismic tomography across the fault zone

The results of the P-wave first arrivals seismic tomography are presented in Figs.11-12. The P-wave active seismic survey was conducted along a ~90 m long profile (t-t' Figs.2, 11a-b) running across the principal fault surface from the hanging-wall into the footwall block. The seismic line laterally overlapped with the gully along which the sampling for the ultrasonic measurements was performed (Fig.11a, b). In particular, the fault zone section exposed within the gully displays a structural setting that can be confidently projected under the seismic line (Fig.11b). The geological cross section along the seismic line (Fig.12a) has been solely drawn on the basis of the VCFZ structural map (Figs.1-2, 12a) and used as an independent comparison with the tomographic model.
Seismic tomographic inversions inherently tend to smooth velocity values in space. Therefore, velocity variations result in sharp gradients in the case of abrupt changes in elastic properties within the rock volume or across major fault surfaces. The well-sampled portion of the P-wave tomographic model was in the range 10-80 m along the acquisition line and reached depths of 15-20 m (Figs.12b). This portion of the model was characterized by very high seismic ray densities (Fig.S5) and a lateral resolution of 2-4 m in the first 10-15 m under the surface (see section S3 for a discussion on the spatial resolution). The less resolved portions of the model (areas in transparency in Figs.12b) have been still used, however, to compare with the internal structure of the VCFZ (Fig.12a; Fondriest et al., 2020).
The seismic P-wave velocities range between 0.4 $kms^{-1}$ and 3.4 $kms^{-1}$ (Fig.12b). The shallowest part of the tomographic model (up to 3-4 m depth) is an almost continuous and homogeneous layer with velocities below 0.9 $kms^{-1}$, corresponding to a regolith covering the fractured bedrock (see the good match with the grassed areas along the profile t-t' in Fig.11a-b). In the central and upper parts of the model, velocity values of 1.2-1.5 $kms^{-1}$ are observed nearby the surface (Fig.12b) and clear lateral velocity anomalies are present at depths greater than 4-5 m (right portion in Fig.12b). The size and geometry of these anomalies are well matched with the position of the different VCFZ footwall structural units (compare Figs.12a and b) and the orientation of both the principal and subsidiary synthetic fault surfaces dipping 50°-65° to the south-west (Fig.12a-b). In the footwall of the principal fault surface a high-velocity body ($V_p$ = 2-3.1 $kms^{-1}$) matches the position of the VCFZ cataclastic fault core (CU1 and CU2; Fig.12a-b). Toward the upper end of the seismic line, another high-velocity body ($V_p$ = 2.5-3.4 $kms^{-1}$) bounded by a sharp interface dipping ~65° to the south-west matches the position of the low-strain damage zone (LSDZ, Figs.12a-b). The zone in between is a ~15 m thick low-velocity layer ($V_p \leq$ 1-1.5 $kms^{-1}$) matching the position of the high-strain damage zone (HSDZ) and corresponding to a higher density of subsidiary faults in the map of Fig.11a. Here, the spatial distribution of the P-wave velocities suggests the occurrence of subsidiary synthetic and possibly antithetic fault surfaces and a south-dipping low-angle feature, well matching with the expected position of the BU and the Omo Morto thrust zone (see section in Fig.12a). Moving to the lower part of the model (left portion in Fig.12b), the hangingwall block of the principal fault surface shows P-wave velocities in the range 1-1.5 $kms^{-1}$, with a body of high relative velocity ($V_p$ > 2 $kms^{-1}$) at depth between 10-15 m from the beginning of the profile (Fig.12b).

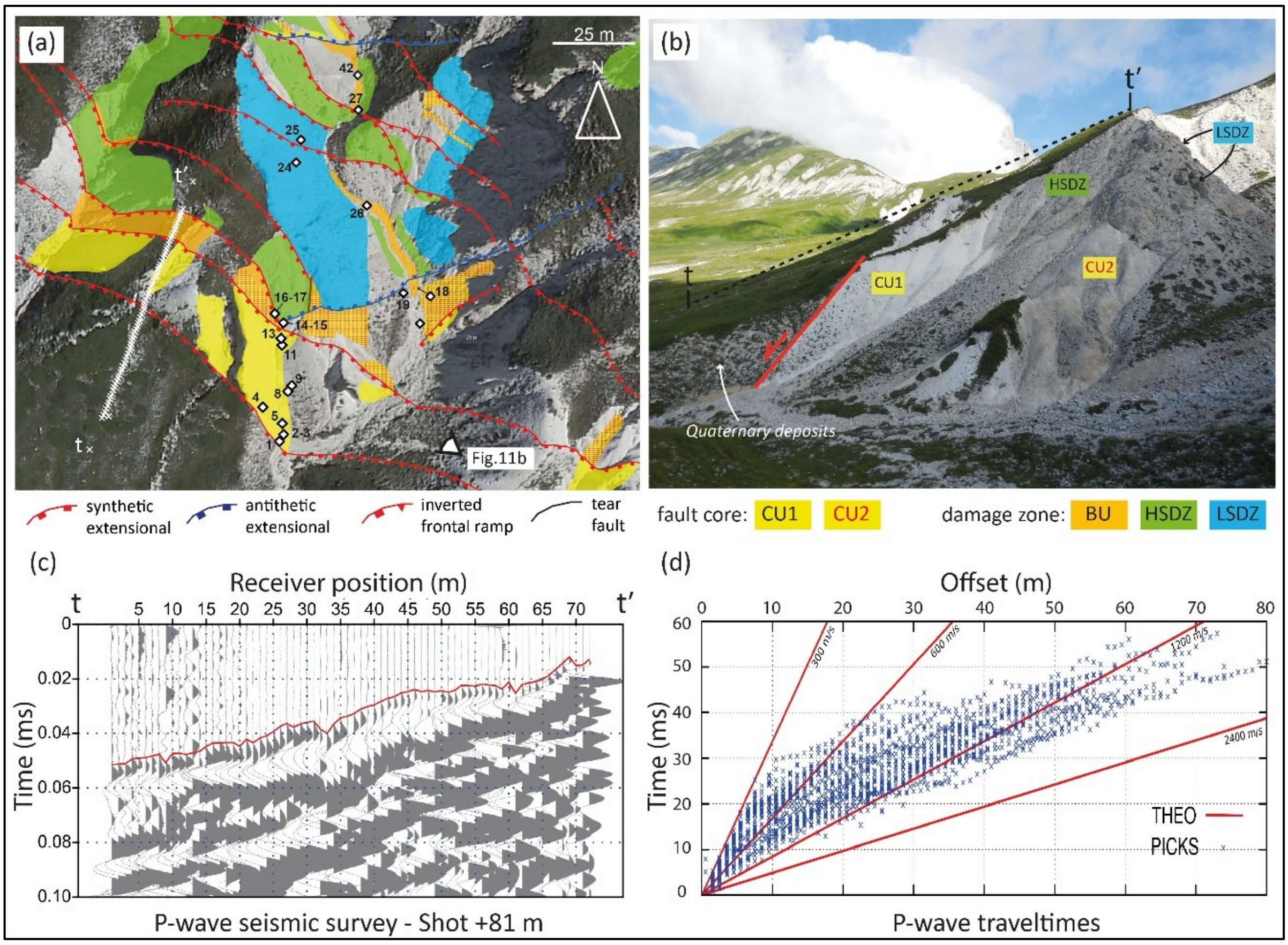

**Fig.11** P-wave first-arrival seismic tomography across the VCFZ. (a) Detail of the structural map of the VCFZ (see Fig.2) in the area where the active seismic profile t-t' has been performed. The seismic profile runs parallel to the creek in which the sampling for ultrasonic velocity measurements has been carried out. (b) Natural section across the VCFZ moving from the hanging-wall (Quaternary deposits to the SW) into the footwall (toward the NE). The main structural units and the trace of the seismic profile are indicated. The structural setting mapped within this creek can be confidently projected under the seismic profile. (c) Seismic sections with common shot (shot at 81 m – top part of the section) registered along the seismic line. The first 0.1 s of registration are displayed in each trace and normalized to the maximum amplitude. Travel-times on y-axis and receiver position on x-axis. The red line identifies the manually picked P-wave first arrivals. (d) P-wave first arrivals (blue crosses) versus offset (i.e. distance between shot and receiver). Red lines show the theoretical travel-times for a uniform propagation medium with different P-wave velocities. P wave seismic velocities are expressed in $ms^{-1}$.

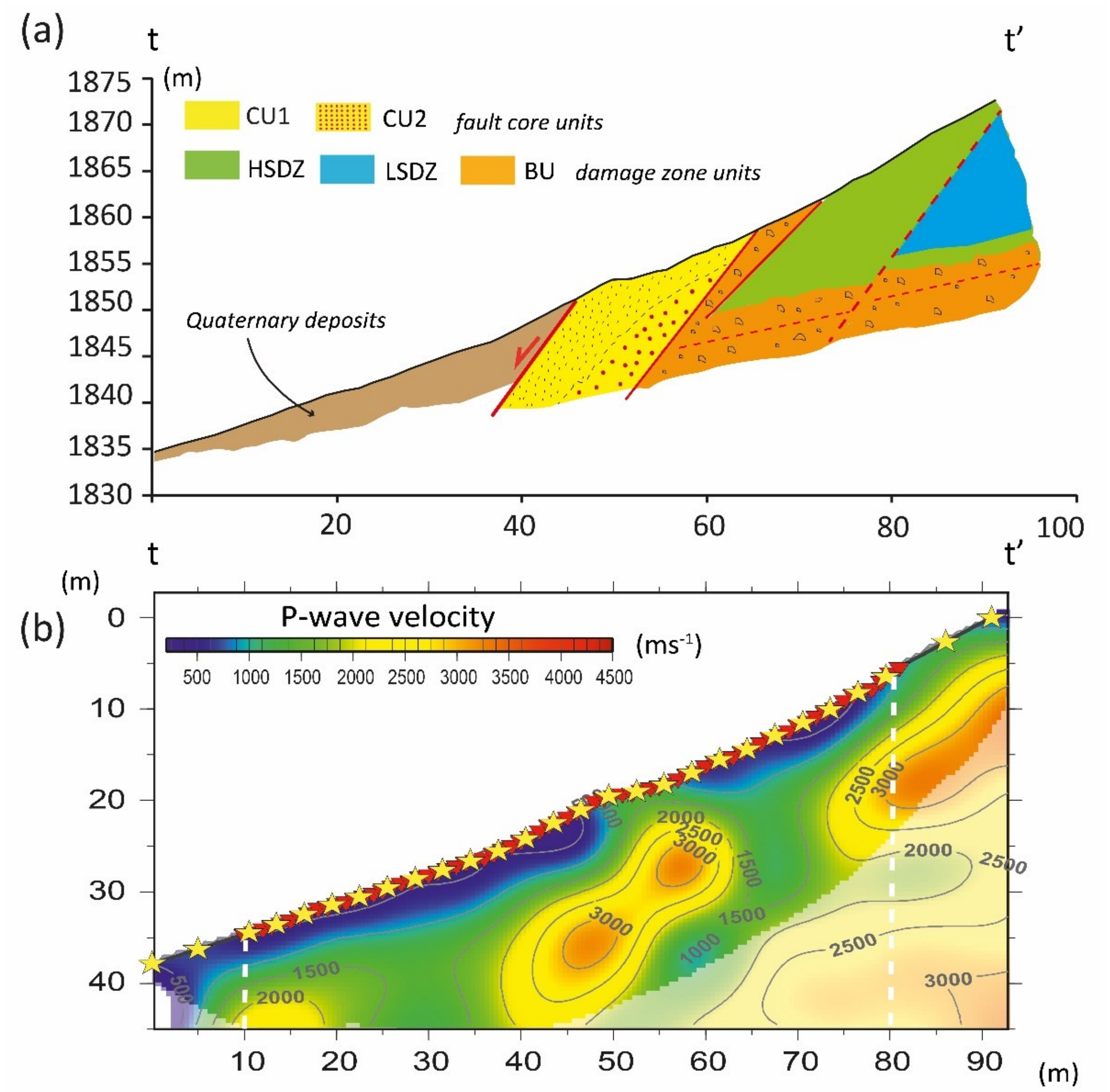

**Fig.12** Vado di Corno fault zone P-wave seismic velocity structure. (a) Structural geological section along the seismic profile t-t'. The section was built and interpreted based only on the structural map of Figs.1-2 and used as an independent comparison with the tomographic model. The legend is the same of Figs.1-2. In the footwall of the principal fault surface, the cataclastic fault core (CU1-CU2, ~ 15 m total thickness) is in contact by fault with the extensional damage zone (HSDZ and LSZD). The low-angle breccia unit (BU) represents the older Omo Morto thrust zone within the extensional damage zone. (b) P-wave final tomographic model obtained from the inversion of first-arrival times (distribution of seismic rays in Fig.S5). On the topographic surface, yellow stars indicate the location of the shots and red triangles the location of the receivers. The best resolved region is between the two white dashed lines and is bounded at depth by the region with overlaid transparency (see section 3.5. for details). The bottom less resolved portion of the model (area in transparency) has been still used, however, to compare the retrieved geometries with the structural geological section t-t'. Areas with contrasting seismic velocities match the size and geometry of the different VCFZ footwall structural units. The high-velocity body (Vp = 2-3.1 kms$^{-1}$) in the central portion of the model matches the position of the VCFZ cataclastic fault core (CU1 and CU2) and shows fault-like sharp boundaries with orientation comparable to that of principal and subsidiary fault surfaces. The high-velocity body (Vp = 2.5-3.4 kms$^{-1}$) in the upper end of the seismic line matches the position of the low-strain damage zone (LSDZ), while the ~15 m thick low-velocity layer (Vp ≤ 1-1.5 kms$^{-1}$) in between corresponds to the HSDZ. The low-angle splaying of such low-velocity layer

well matches with the expected position of the BU and the Omo Morto thrust zone. See section 4.4. of the main text for a detailed description. P wave seismic velocities in *panel b* are in $ms^{-1}$.

### 4.5. Mesoscale fracture intensities across the fault zone

Areal fracture intensity P21 ($m^{-1}$) has been determined for the fault core units CU1-CU2 and for the extensional damage zone units HSDZ and LSDZ at seven locations (Table 3, Figs.S9-11). On the contrary, P21 was not determined for the breccia unit (BU) due to its pellicular exposure across the fault zone and because, depending on the location, it is part of the high- or the low-strain damage zone.

Areal fracture intensity was 22-50 $m^{-1}$ for the fault core units (CU1-CU2), 111-143 $m^{-1}$ for the HSDZ and 8.49-8.82 $m^{-1}$ for the LSDZ (Table 3). The fracture size distributions were also different, with the lowest slopes (D) for the LZDZ, intermediated slopes for the HSDZ and the highest slopes for CU1-CU2 (Table 3, Figs.13 and S12). Such data indicate that, at the mesoscale, the HSDZ outcrops contain more open fractures than the other structural units, while the LSDZ show both the lowest fracture intensity and smallest amount of short fractures (Fig.12). It is worth to remind that we counted as fractures: faults, shear fractures with negligible slip, joints, bedding surfaces. On the contrary sealed microfractures and veins were not considered (see section 3.6.).

| Outcrop | Structural unit | Orientation | Range (m) | Scan area ($m^2$) | P21 ($m^{-1}$) | D |
|---|---|---|---|---|---|---|
| CU-1 | CU1 | fault Xh | 0.042-0.459 | 0.208 | 22.41 | -1.59 |
| CU-2 | CU2 | fault Xh | 0.02-0.103 | 0.123 | 49.58 | -1.74 |
| HSDZ-1 | HSDZ | fault Xv | 0.009-0.182 | 0.220 | 142.91 | -1.55 |
| HSDZ-2 | HSDZ | fault //v | 0.01-0.182 | 0.239 | 129.79 | -1.57 |
| HSDZ-3 | HSDZ | fault Xv | 0.006-0.308 | 0.240 | 111.83 | -1.53 |
| LSDZ-1 | LSDZ | fault Xv | 0.04-0.124 | 0.673 | 8.82 | -1.37 |
| LSDZ-2 | LSDZ | fault Xv | 0.03-0.178 | 0.281 | 8.49 | -1.17 |

**Table 3.** Data of mesoscale fracture characterization of the VCFZ structural units (CU, HSDZ, LSDZ). Progressive numbering indicates the outcrop where the measurements have been performed (outcrop locations are in the data repository). Orientation of the outcrop exposures is indicated as: fault perpendicular horizontal (fault Xh, pavement exposure), fault perpendicular vertical (fault Xv, section exposure) and fault parallel vertical (fault //v, section exposure). Range (m) is the range over which the crack size distribution is linear and D has been calculated. P21 is the areal fracture intensity calculated on the same window.

**Fig.13** (a) Examples of mesoscale fracture tracing for the VCFZ structural units CU, HSDZ and LSDZ. The units showed different fracture density and size distributions at length scales up to 1-1.2 m. The fracture size distributions were described by power law LOG N = -D LOG r + C in their linear portion, being N the cumulative number of fractures, r the fracture half-length and D the slope of the line. The effective medium approach of Budiansky and O'Connell (1976) was applied iteratively by dividing the fracture size distributions in small bins with fracture density values << 0.4. See sections 6.2. of the main text for a detailed description of the theory and the analysis performed. See the Supporting Information for the entire dataset of fracture size distribution (Figs.S9-12).

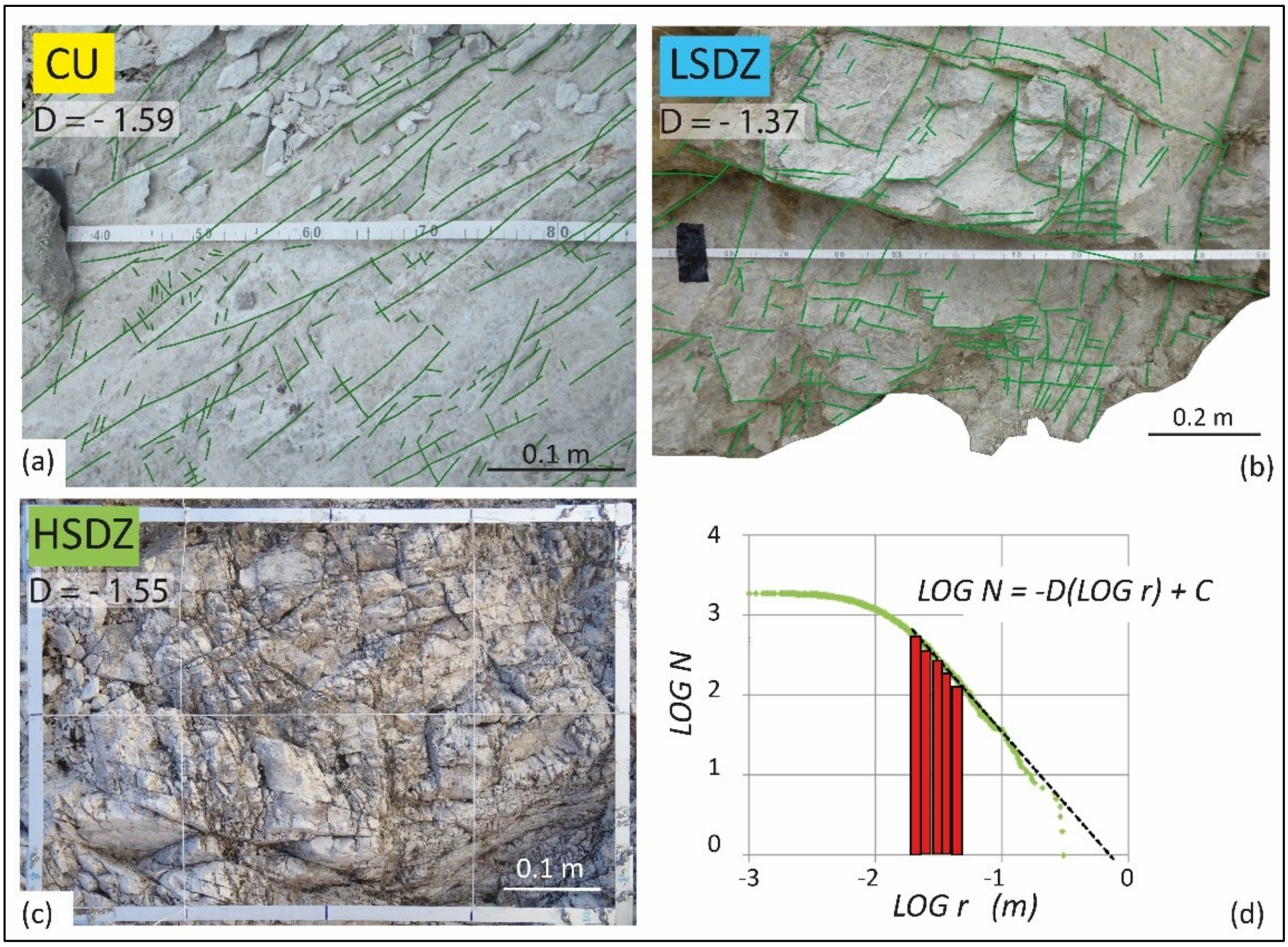

**5. Discussion**

**5.1. Petrophysical control on the ultrasonic velocity structure of the VCFZ**

The ultrasonic velocity structure of the VCFZ was determined through laboratory measurements on centimetre-sized fault zone rock samples. As a main outcome, a gradual increase in ultrasonic wave velocities with increasing distance from the principal fault surface was not observed at this scale (Fig.6). Instead the fault zone was characterized by the juxtaposition of structural units with sharply distinct ranges of ultrasonic velocity (Fig.6). In particular, the fault core cataclastic units, CU1 and CU2, were significantly “slower” compared to the damage zone units, HSDZ and LSDZ (on average by about 1.5 $kms^{-1}$ for P- and 0.7 $kms^{-1}$ for S-wave). The breccia unit (BU), which is the expression of the older Omo Morto Thrust Zone (Vezzani et al., 2010) outcropping within the extensional damage zone, displayed intermediate velocities. Directional ultrasonic measurements pointed out an elastic anisotropy <15-20% for P-waves that varies moderately across the fault zone structural units. The lowest speed was in the NNE-SSW direction orthogonal to the strike of the main extensional faults and fracture sets (45% of all the samples), while about 30% of the samples were slower in the fault parallel horizontal direction (Fig.6c). Samples from the low-strain damage zone (blue unit Figs.2-4) were slower in the fault vertical direction being affected by the presence of a low-angle bed-parallel layering (Fig.10d).

Both average and directional wave velocities showed a negative correlation with total connected porosities, described fairly well by exponential decays (Fig.7, Fig.14, Fig.S6-8). However, the presence of a large data scattering and clear outliers (e.g. *samples 29 and 32*) with respect to these trends revealed a significant samples variability in some of the structural units, especially within the fault core (CU1-CU2) and the high-strain damage zone (HSDZ) (Figs.6b, 7). This point was supported by the pore-size distributions and the microstructural observations (Figs.

8-10) highlighting the fundamental control of pore types and rock microstructure on ultrasonic velocities.

It is well known that acoustic properties in carbonate rocks are controlled by the combination of depositional and post-depositional fabrics, the latter produced by the early activation of diagenetic processes, such as dissolution and cementation, during compaction. This implies that, unlike for siliciclastic rocks, carbonates show little correlation of seismic velocities with age or burial depth of the sediment, and velocity inversions with depth are common (Anselmetti and Eberli, 1993). In addition, carbonate rocks have heterogenous pore networks characterized by different pore types and sizes, resulting in complex velocity-porosity relations often associated with a large scatter of ultrasonic velocities for a given porosity (Anselmetti and Eberli, 1997). The pore types mainly include interparticle or intercrystal pores, moldic pores, intraparticle pores, vugs and microfractures (Lucia, 1983). In general, intraframe pores (i.e. molds and intraparticle) can result in high velocities even for high porosity values, while microporosity (i.e. interparticle, intercrystal, microfractures) is associated with much lower velocities at the same porosity values (Anselmetti and Eberli, 1997; Baechle et al., 2004; Bakhorji and Schmitt, 2022). Several elastic models have been developed to account for such pore network complexity in carbonates. For instance, Saleh and Castagna (2004) described carbonates porosity as a mixture of interparticle/intercrystalline and near-spherical pores. According to this model, rocks with only interparticle/intercrystalline porosity will follow the Willie time-average equation trend, while velocities would increase toward a maximum limit when all pores are spherical or near spherical. Weger et al. (2004) further suggested that, at a given porosity value, acoustic properties are variably influenced by three pore shape parameters: pore roundness, dominant pore size and tortuosity. Tortuosity (ratio of perimeter to area - PoA) was the most relevant parameter (i.e. velocities are higher for less tortuous pore systems) while the pore roundness was the least influential one (Weger et al., 2004). Lastly, Xu and Paine (2009) proposed a carbonate rock physics model predicting both positive velocity variations from the time-average reference trend due to the presence of stiff rounded pores and negative ones due to the occurrence of compliant microfracture porosity both in limestone and dolostones (see also Wang et al., 2015).

Similar concepts have been more recently used to explain ultrasonic wave velocities in carbonate-bearing fault rocks exhumed from shallow depth (<2 km) and their relation with type/distribution of pores, intensity of rock deformation and fluid-rock interaction (i.e. structural diagenesis; Ferraro et al., 2020; Laubach et al., 2010; Matonti et al., 2015). In Fig.14 we plot the average P-wave velocity data *versus* He-porosities for all the analyzed rock samples of the VCFZ. The data are plotted with respect to the Hashin–Shtrikman upper and lower bounds ($HS^+$ and $HS^-$) of calcite and dolomite, the time-average equation for calcite- and dolomite-air system (Wyllie et al., 1956, 1958, 1963), velocities of non-porous calcite and dolomite, and the best-fit curve published by Anselmetti and Eberli (1993) for the platform carbonates of Bahamas and Maiella Mt. (Italy). All data plot below the best-fit line of Anselmetti and Eberli (1993) that actually grouped measurements of dry and water-saturated carbonate samples under an effective pressure of 8 MPa. Only *sample 35* from the HSDZ (porosity of 0.35±0.04 %) plotted on the Hashin–Shtrikman bound of calcite, considering that it consists almost entirely of dolomite with some small vugs and secondary dolomite veins. Samples with comparable porosities lower than 2% are clustered in two groups exhibiting very different velocities (Fig.14). The first group shows P-wave velocity larger than 5.5 $kms^{-1}$ and includes samples from the high- and low-strain damage zone at most affected by few microfractures and sealed dolomite veins. In the case of the HSDZ samples, dolomitization is intense and can partially or totally obliterate the original sedimentary fabric (e.g. Fig.10b), while in the LSDZ lens the intraclasts-rich sedimentary layering is well preserved (e.g. Fig.10d). These samples follow fairly well the time-average trend for the dolomite-air system, with a dominant interparticle or intercrystal porosity. The occurrence of some molds and intraparticle pores, more frequent in the LSDZ, can explain the positive variations from this trend (Fig.14). The second group shows P-wave velocity lower than 5.0 $kms^{-1}$ and comprises samples of HSDZ, BU and CU2, all lying below the dry dolomite time-average curve (Fig.14). Here the largely reduced velocities are controlled by a compliant microfracture-dominated porosity. Both the HSDZ outliers (*samples 29 and 32*), some BU and CU2 samples are affected by a network of fault parallel and fault perpendicular shear fractures related to the extensional deformation phase and associated with localized cataclasis and only partial sealing by calcite cements (e.g. Fig.10a). As suggested by several authors, the sealing of fractures by cements can change and even cancel the way in which fractures affect waves propagation and their speed (e.g. Matonti et al., 2015; Vanorio and Mavko, 2011). Indeed, samples affected by fractures completely sealed by dolomite sparite (i.e. thrust-related veins in the HSDZ and BU) have the highest wave speeds, while rock comminution along new or reactivated fractures (HSDZ outliers, BU, CU2) has a strong effect in lowering ultrasonic wave speeds (see also results from Acosta-Colon et al., 2009; Fortin et al., 2007; Matonti et al., 2017).
Samples from the proximal fault core (CU1 unit) display completely different velocity-porosity relations suggesting a strong microstructural effect (Fig.14). Indeed, P-wave velocities are highly scattered and samples with very different porosity values show comparable wave speeds (Fig.14). We interpreted these data as a result of the complex relation between evolving textural maturity of the fault rocks and the degree of pore space sealing by calcite cement (Figs.9, 14). Samples with a protocataclastic fabric ("pc" in Fig.14; Fig.9a-b) follow the time-average trend and connected porosity is both intergranular and along fractures (intra- and transgranular ones). Samples with a cataclastic fabric ("c" in Fig.14) are characterized by rounded clasts surrounded by a larger amount of fine-grained and virtually compliant dolomite matrix with many sub-micrometre pores (Figs.8, 9c). Samples with a ultracataclastic fabrics ("uc" in Fig.14) represent the extreme case with matrix-supported fabrics and higher porosity. However, both cataclasites and ultracataclasites plot above the time-average trend and their final ultrasonic velocities depend on the degree of pore space sealing by calcite cement associated with cyclic fracturing and fluid infiltration during the extensional activity of the VCFZ (see Agosta and Kirschner, 2003; Demurtas et al., 2016; Lucca et al., 2019). Indeed, the progressive reduction of the connected pore network by calcite cement leads to the generation of more isolated sub-rounded micro-pores and results in stiffer fault rock frameworks and relatively high ultrasonic velocities.

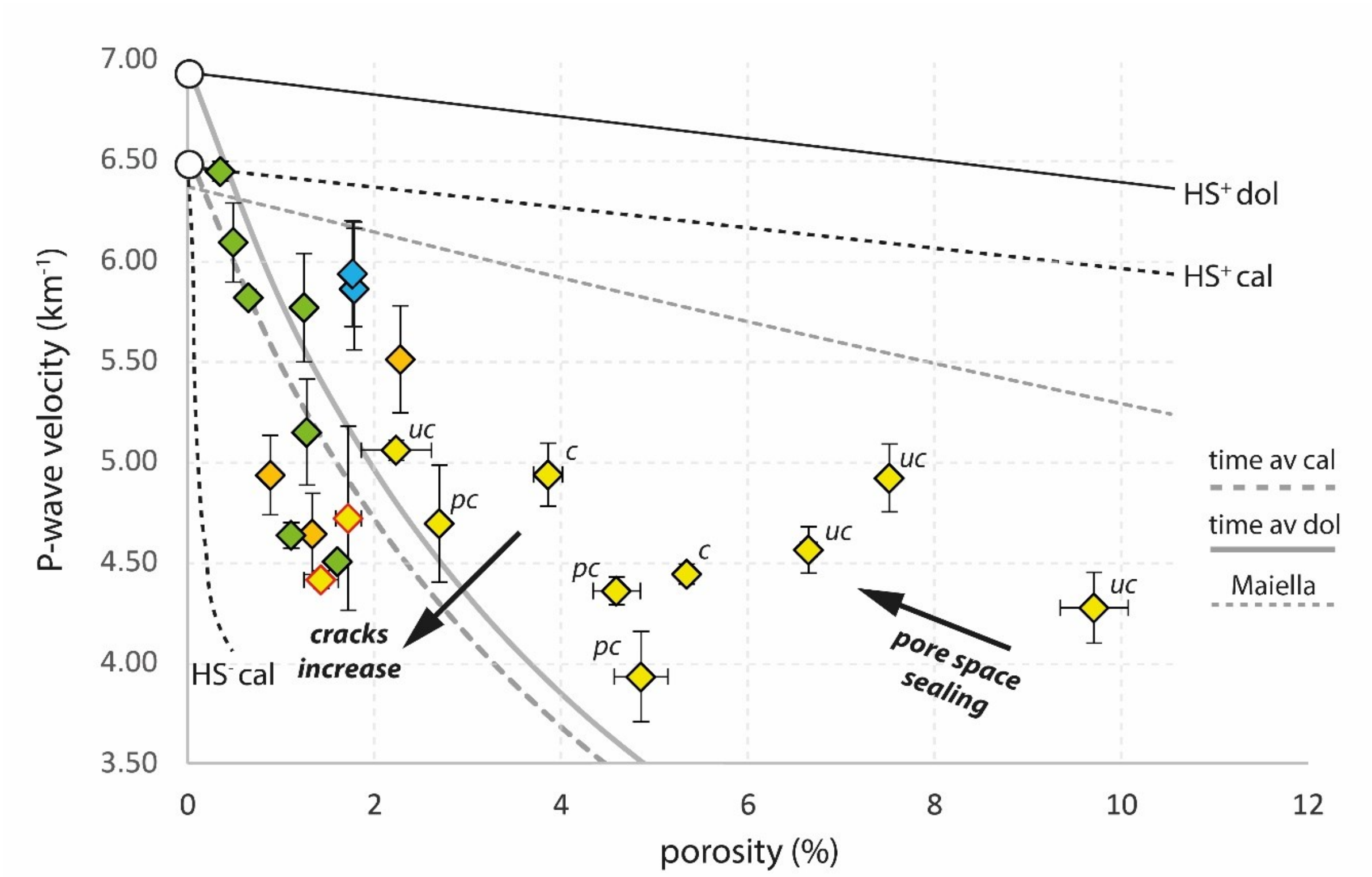

**Fig.14** Petrophysical control on the ultrasonic velocities of the VCFZ. The plot displays the dependence of P-wave ultrasonic velocities versus He-porosities for the analysed fault zone rock samples. Data are plotted against the Hashin–Shtrikman upper and lower bounds (HS+ and HS -) of calcite and dolomite, the time-average equation for calcite- and dolomite-air system (Wyllie et al., 1956), velocities of non-porous calcite and dolomite, and the best-fit curve published by Anselmetti and Eberli (1993) for the platform carbonates of Bahamas and Maiella Mt. (Italy). See section 5.1. for a detailed description.

**5.2. High-resolution seismic velocity structure and field-scale heterogeneity of the VCFZ**

The high-resolution P-wave first arrival seismic tomography showed strong lateral variations of seismic velocities across the VCFZ footwall, below a discontinuous layer of low-velocity regolith, with a maximum thickness of 4-5 m (Figs.12, 15a). Given the inherent tendency of seismic tomography to smooth out velocity variations, the strong lateral velocity gradients observed in the tomographic model clearly indicate the occurrence of rock volumes with very contrasted elastic properties and sharp boundaries (Figs.12, 15a). In particular, both the dimensions of these bodies (thickness ≥ 10 m) and their geometry nicely compare with those of the fault zone structural units mapped in the field (Fig.12a, b; Demurtas et al., 2016; Fondriest et al., 2020). The lateral contacts are well matched to the position and orientation of the principal and subsidiary faults, both synthetic and antithetic, exposed in the outcrops (Fig.15a). In contrast, no increase of seismic velocity is observed with depth within the different elastic bodies (i.e. structural units) of the VCFZ footwall (Fig.12b, 15a).

The high-velocity body in the central portion of the tomographic model corresponds to the fault core cataclastic units (CU1 and CU2), while the one in the upper part of the model (right side in Figs.12, 15a) corresponds to the low-strain damage zone (LSDZ). Internal lenses with P-wave velocity ≥ 3 $kms^{-1}$ are present within the fault core, while velocities grade down to ~2 $kms^{-1}$ at lateral boundary faults. This is the case for the LSDZ southern boundary as well (Figs.12, 15a). In between the relatively fast fault core (CU1-CU2) and low-strain damage zone (LSDZ) units, a 10-15 m thick layer with P-wave velocity as low as ~1 $kms^{-1}$ matched well the position of the high-strain damage zone (HSDZ). The HSDZ position also coincides with the presence of densely spaced macro-scale faults, interacting and cross-cutting each other (Figs.11a, 12, 15a). This low-velocity layer branches out along a low-angle body dipping to the south that matches well with the setting of the Omo Morto thrust zone and the breccia unit (BU). The predicted positions of the principal fault surface of the extensional VCFZ and the older thrust are in good agreement with those inferred from the seismic tomographic model (Figs. 12 and 15a). On the other side, the hangingwall block of the VCFZ consists of a unit with P-wave velocity in the range 1-1.5 $kms^{-1}$, but a relative higher velocity body ($V_p > 2$ $kms^{-1}$) emerges at ~10 m depth in the southern part of the section (Figs.12b, 15a). Since there are no continuous outcrops allowing to determine the structure of the hangingwall block at depth, we relied on few discontinuous surface information. Based on that, the hangingwall block consists of colluvial gravel-like Quaternary deposits with local cohesive lenses. Therefore, the higher velocity body at the base of the modelled section may represent a stiff heterogeneity within the Quaternary deposits or another fault core (along a southern strand of the fault) developed either in bedrock or in cemented Quaternary deposits. The latter option would imply that another tectonic sliver is present in the hanging-wall of the mapped principal fault surface, although the tomographic model is not well resolved in this portion.

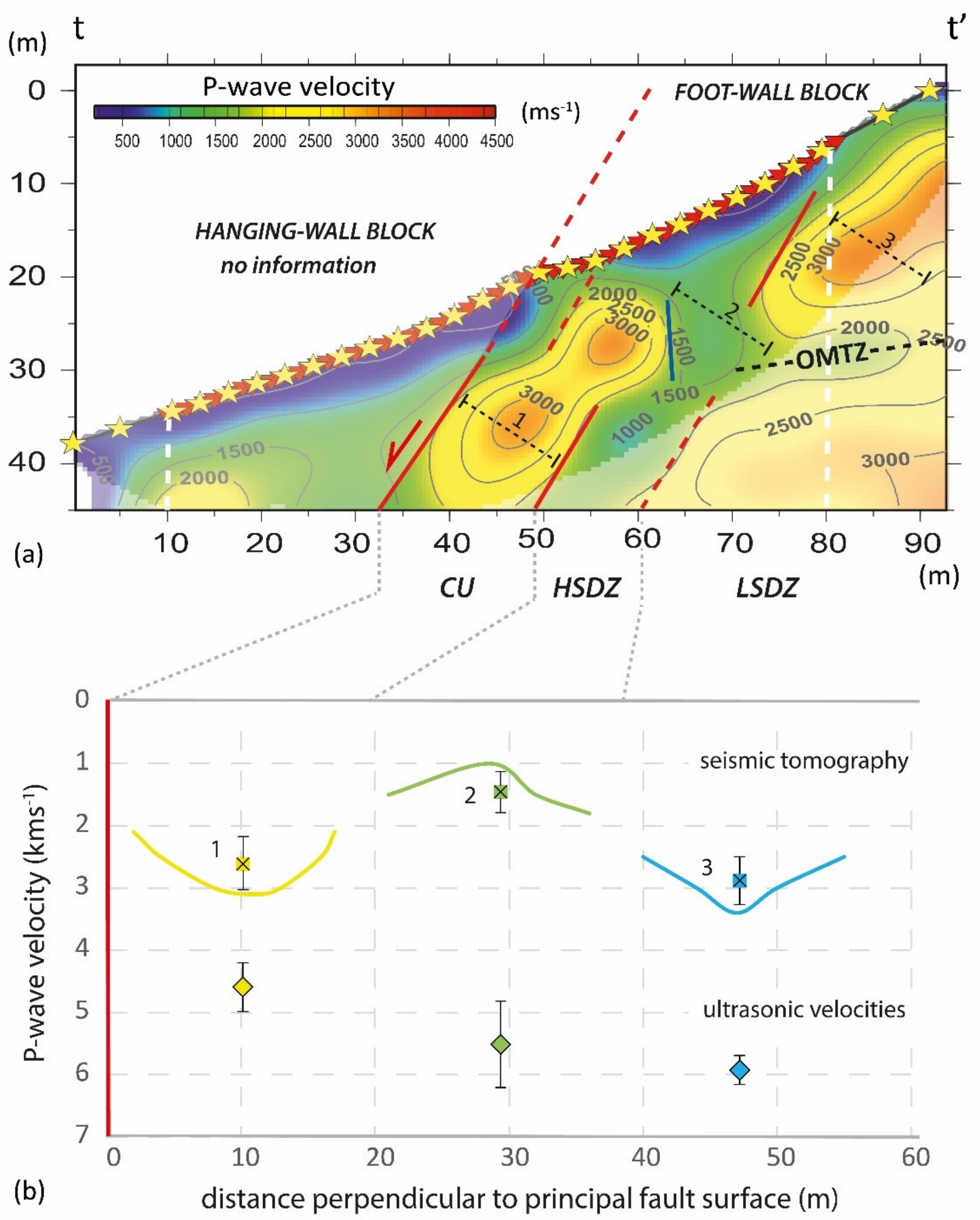

**Fig.15** Seismic velocity structure and scaling in the VCFZ. (Upper panel) Vado di Corno fault zone P-wave seismic velocity structure. Areas in transparency are the hanging-wall block and the bottom part of the footwall. The contacts among the different structural and velocity units are sharp and match the orientation of the principal and subsidiary faults, both synthetic (red) and antithetic (blue). The low-angle structure of the Omo Morto thrust zone is also well recognizable. Dashed lines are inferred contacts. Profile 1-3 (thin dashed black lines) have been drawn through the fault core cataclastic units (CU, profile 1), the HSDZ (profile 2) and the LSDZ (profile 3) in direction perpendicular to the principal fault surface. (Bottom panel) The P-wave seismic velocities along profiles 1-3 together with the average values are reported for the structural units CU, HSDZ and LSDZ. As a comparison, the average ultrasonic velocities for the same structural units are also reported. There is a distinct velocity scaling for each different structural unit. See section 5.2. of the main text for more details. P wave seismic velocities in *panel a* are in $ms^{-1}$.

### 5.3. Scaling of the VCFZ velocity structure through an effective medium approach: from the laboratory to the field scale

The fault zone rock samples collected in the VCFZ display ultrasonic velocities that are always higher compared to the seismic velocities derived from the P-wave tomography model (Fig.15b). Differences between laboratory measurements of ultrasonic wave speeds and those determined in the field by active seismic surveys have been reported in literature and possibly related to selective rock sampling, effects of fluid saturation, or frequency dependence of elastic properties due to heterogeneity and/or anisotropy varying with the investigated length scale (e.g. Rempe et al., 2013; Stierman and Kovach, 1979; Taylor et al., 2015). In the case of the VCFZ, the rock sampling was performed avoiding any mesoscale fault and fracture so that the measured ultrasonic wave speeds were representative of the bulk elastic properties of the "matrix". Changes of elastic wave speeds cannot be ascribed to fluid saturation effects since the study area is a dry karstic region dominated by intensely fractured carbonates and the survey was conducted during the dry summer season. On the other hand, based on our field observations, the reduction of wave speeds from the laboratory to the field scale is most likely controlled by the dispersion and attenuation of elastic waves due to the effect of mesoscale fractures. In addition, the magnitude of the velocity reductions from the laboratory to the field scale varied among the different structural units (Fig.15b). At the laboratory scale the fault core cataclastic units (CU1 and CU2) are slower compared to the high- and low-strain damage zone units (HSDZ and LSDZ). In turn, at the scale of the seismic survey, the cataclastic fault core units and the low-strain damage zone appear as relatively fast and rigid bodies compared to the very slow and compliant heavily fractured high-strain damage zone (Fig.15b).
Hence, we attempt to model the reduction of elastic wave speeds from the length scale of the ultrasonic laboratory measurements (0.04 m of average sample size) to that of the seismic tomography (2-4 m of estimated maximum horizontal resolution, see section S3) through a simplified effective medium approach (Taylor et al., 2015) considering the effect of mesoscale fracture density and size distribution in each structural unit (Figs.13, 16). Various effective-medium theories exist relating the elastic moduli of a fractured rock to those of a fracture-free rock and to fracture density (Jaeger et al., 2007, Fig.S13). The self-consistent effective medium approach of Budiansky and O'Connell (1976) was adopted which offered easy formulations under the assumption of an isotropic population of uniformly-sized penny-shaped fractures:

$$G_C = G\left(1 - \frac{32}{45}\,\frac{(1-\nu_C)\,(5-\nu_C)}{(2-\nu_C)}\,\varepsilon\right) \qquad (1)$$

$$K_C = K\left(1 - \frac{16}{9}\frac{(1-\nu_C{}^2)}{(1-2\nu_C)}\,\varepsilon\right) \qquad (2)$$

$$\nu_C = \nu\left(1 - \frac{16}{9}\,\varepsilon\right) \qquad (3)$$

where *G, K, v* are respectively shear modulus, bulk modulus, Poisson's ratio of the intact rock, while the subscription *c* refers to the elastic properties of the fractured medium. The term $\varepsilon$ is defined as the *crack density parameter* and represents an approximate 3D fracture density estimated from two-dimensional fracture traces (see section 3.6.) following the expression by Budiansky and O'Connell (1976):

$$\varepsilon = \frac{8}{\pi^3}\,M < r^2 > \qquad (4)$$

where *M* is the number of fractures per unit area in 2D and $< r >$ is the average half-length of those fractures. This approximation is equivalent to $\varepsilon \equiv Nr^3/Vol$ for a population of *N* similarly sized fractures per unit volume *Vol* (Budiansky and O'Connell, 1976; Taylor et al., 2015). The self-consistent method of Budiansky and O'Connell (1976) (equations 1-3) would predict unrealistic elastic moduli for $\varepsilon$ larger than 0.4 (linearly decreasing to zero at $\varepsilon$ = 0.56), but provides results comparable with other more sophisticated effective medium theories (e.g. differential or non-interaction assumption; Gueguen and Kachanov, 2011; Kachanov, 1993) for very low fracture densities (Fig.S13). Based on this argument, the self-consistent method (valid for uniformly-sized fractures) was applied by discretizing the fracture size distributions of each structural unit in fracture size bins smaller enough to lead values of $\varepsilon$ << 0.4 (Fig.13). By doing so, the progressive reduction of elastic moduli was calculated starting from the elastic moduli values determined at the average laboratory length scale of 0.04 m to that of the seismic tomography models of 2-4 m (Fig.16), and consequently the evolution of P- wave velocities following Jaeger et al. (2007):

$$V_{Pc} = \sqrt{\frac{K_C+\frac{4}{3}G_C}{\rho}} \qquad (5)$$

where $V_{PC}$, and $\rho$ are respectively effective velocity of P- wave and bulk density.
The fracture size distributions were determined for the fault core units CU1-CU2 and for the extensional damage zone units HSDZ and LSDZ over scan areas ranging from 60x40 cm$^2$ to 80x100 cm$^2$ (see sections 3.6. and 4.5. for details). The power-law parameters describing each distribution are reported in Table 3 and Figs. 13 and 16, while the single plots are in Fig.S12. The results of the modelling are displayed in Fig.16. To account for spatial variability effects, we report the modelling based on the 2D fracture size distributions determined on two different scan areas for each structural unit (Fig.16). Given that the dominant contribution to seismic velocities in the tomography model should come from horizontal ray paths, we consider outcrop-scale fracture size distributions determined on sections perpendicular to the principal fault surface.
As a result of the self-consistent effective medium modelling the fault core units CU1 and CU2 (treated together; yellow curves in Fig.16) evolve from an initial average P-wave velocity value of ~4.8 kms$^{-1}$ at the laboratory length scale (0.04 m) to ~2.8 kms$^{-1}$ for a maximum fracture length of 2 m. The LSDZ (light blue curves in Fig.16) P-wave velocities evolve from ~6 kms$^{-1}$ to ~3.4 kms$^{-1}$, while the HSDZ ones rapidly drop from ~5.5 kms$^{-1}$ to 1.1 kms$^{-1}$ for the same maximum fracture length. The BU data are not represented since, as part of the VCFZ extensional damage zone (HSDZ or LSDZ), they would follow similar trends to these units but starting from lower initial laboratory scale velocities. Interestingly the final modelled P-wave velocities are at first-order comparable with those derived from the seismic tomography, suggesting that seismic velocity values of the VCFZ structural units are mainly controlled by the fracture density and the fracture size distributions at the length scale of few meters. Here, fractures comprise faults dominating in the fault core units, joints and bedding surfaces dominating in the damage zone units (see section 3.6. for more details). Given the unrealistic assumption of non-interaction of fractures behind the modelling, the resulting P-wave velocities need to be considered as upper bound values. It is also worth to note that these velocity estimates are affected by uncertainties due to the fact that the fracture size distributions have been extrapolated to 2 m length scales (using the slopes *D* determined up to a maximum length scale of 1.2 m) and the assumption of an isotropic distribution of fractures underlying the calculations of 3D fracture densities from 2D data (additional data from fault parallel and fault perpendicular sections were also considered to minimize this source of error, see Fig.S12).
However, it is important to stress that such a simple modelling approach, which provides a first-order approximation of the effect of mesoscale fracture on seismic velocities scaling, was successful in matching the velocity ranges obtained in the high-resolution P-wave seismic tomography. Moreover, the lateral velocity anomalies recognized in the shallow and higher resolution portion of the tomography model are also recognizable in the deeper portion of the model, while no velocity changes are observed vertically with depth (Figs.12b, 15a). This increased our confidence in the modelled seismic velocity values down to 20-25 m, where the vertical load is comparable to that applied during the laboratory ultrasonic velocity measurements (~0.6-0.8 MPa).

| Structural units | CU1 | CU2 | CU | BU | HSDZ | LSDZ |
|---|---|---|---|---|---|---|
| | n (10) | n (2) | n (12) | n (4) | n (9) | n (4) |
| P-wave vel. (km/s) | 4.59 ± 0.41 | 4.61 ± 0.35 | 4.59 ± 0.39 | 5.06 ± 0.42 | 5.51 ± 0.69 | 5.93 ± 0.23 |
| S-wave vel. (km/s) | 2.65 ± 0.22 | 2.61 ± 0.13 | 2.65 ± 0.21 | 2.70 ± 0.29 | 3.22 ± 0.35 | 3.38 ± 0.09 |
| Dynamic bulk modulus (GPa) | - - | - - | 31.41 ± 2.41 | 44.01 ± 3.52 | 46.42 ± 5.10 | 55.13 ± 1.77 |
| Dynamic shear modulus (GPa) | - - | - - | 18.76 ± 2.09 | 20.17 ± 3.05 | 28.88 ± 4.42 | 31.83 ± 1.25 |
| Dynamic Young mod. (GPa) | - - | - - | 46.94 ± 7.08 | 52.49 ± 9.74 | 71.76 ± 15.14 | 80.08 ± 4.63 |
| Dynamic Poisson ratio | 0.25 ± 0.02 | 0.26 ± 0.02 | 0.25 ± 0.02 | 0.31 ± 0.04 | 0.24 ± 0.02 | 0.22 ± 0.04 |

**Table 4.** Summary of the elastic properties of the VCFZ structural units for the EMT modelling.

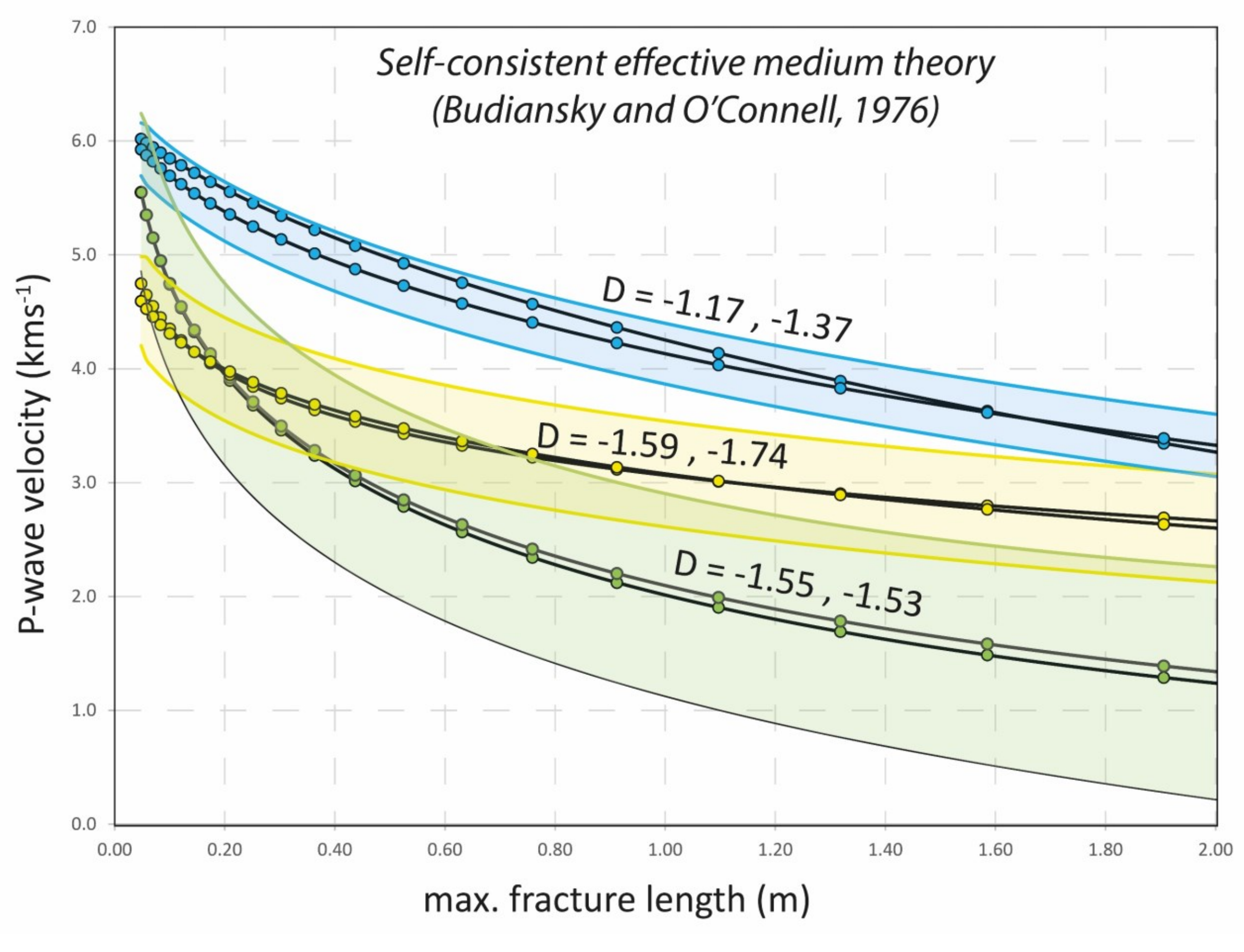

**Fig.16** Evolution of seismic P-wave velocities with increasing maximum fracture length for the structural units CU, HSDZ and LSDZ. The initial velocity values, determined in the laboratory as ultrasonic velocities at the length scale of few centimetres (0.04 m as a reference sample size), were scaled up to length scales comparable with the spatial resolution of the seismic tomography models (2-4 m). The colour code is yellow for the fault core unit (CU), green for the HSDZ and light blue for the LSDZ. The values D refers to the slopes of the fracture size distributions as determined in Figs.13, S12. Each curve refers to a fracture size distribution determined on a different window in the field. The fracture size distributions were determined in fault perpendicular sections. See section 5.3. of the main text for more details.

**Section 5.4. Implications of the heterogeneous seismic velocity structure for near-surface fault zone mechanics**

The high-resolution seismic tomography across the VCFZ highlighted a heterogeneous near-surface velocity structure characterized by ≥10 m thick fault-bounded rock bodies, each with different velocity ranges, that correlate very well with the dimension and position of structural units from previous mapping (Demurtas et al., 2016; Fondriest et al., 2020). In particular, the cataclastic fault core units and the low-strain damage zone appear as relatively "fast" (stiff) bodies in contrast with the very "slow" (compliant) and heavily fractured high-strain damage zone. The majority of previous high-resolution seismic tomography studies in the Central Apennines focused on the characterization of fault-bounded hanging-wall basins and the retrieved seismic velocities mainly refer to Quaternary deposits rather than to bedrock units (e.g. Improta et al., 2007; Maraio et al., Tectonophysics, 2023; Villani et al., 2016, 2021, 2024). On the other side, regional-scale seismic tomographies did not resolve any "low-velocity" fault-related zones in the area (e.g. Chiarabbba et al., 2020; Buttinelli et al., 2018; Di Stefano et al., 2011). Therefore, the seismic tomography imaging across the VCFZ is, to our knowledge, the first high-resolution quantification of the near-surface velocity structure of an exhumed seismogenic fault zone in carbonate rocks. As a consequence, the recognized internal contrasts of seismic velocity are novel observations if compared with the rather homogeneous low velocities detected across fault zones at different scales by previous studies (e.g. Catchings et al., 2014; Mooney and Ginzburg, 1986; Qiu et al., 2018; Rempe et al., 2013; Spudich and Angstman, 1980). Our results are instead more consistent with recent state-of-the-art high-resolution tomographies based on the passive acquisition of both noise and earthquake data at very dense arrays across active fault zones (Roux et al., 2016; Share et al., 2017; Touma et al., 2021, 2022). For instance, these studies highlighted the presence of tens meters thick fault-parallel layers with different seismic velocities (Roux et al., 2016) and small-scale seismic scatterers indicative of a highly heterogeneous internal structure (e.g. Touma et al., 2022).

The VCFZ is characterized by the exposure of the intensely fractured HSDZ (high strain damage zone; see Fig.1b and Fondriest et al., 2020) with thickness up to few hundred meters. This extensive and compliant damage zone (Vp ~1 $kms^{-1}$ at length scale of a few meters) is associated to a geometrical and kinematic complexity due to the interaction of the high-angle extensional VCFZ and the older low-angle Omo Morto thrust zone (Fondriest et al., 2020). Similar thick and intensely fractured damage zones have been also reported along other exhumed and seismogenic fault zones cutting tight carbonates (Fondriest et al., 2015, 2017; Schröckenfuchs et al., 2015; Cortinovis et al., 2024). In general, these fault zones are exhumed from <2 km depth and the rock damage exposed at the surface is the result of a cumulative deformation history plus exhumation. Regardless of the origin of this extensive damage and of the structural level at which it was mostly generated, its occurrence at depths <1 km will likely affect earthquake surface ruptures, the distribution of co-seismic surface deformation and ground motion. For instance, a new-generation of geodetic observations based on correlation of high-resolution aerial imagery before and after large earthquakes, increasingly reported cases in which 40-50% of co-seismic surface deformation occurred off of the principal fault surface within the surrounding medium over thickness up to few hundred meters (e.g. 1992 Landers, 2013 Balochistan, and 2019 Ridgecrest M > 7 mainshocks; Antoine et al., 2021; Milliner et al., 2015; Zinke et al., 2014). The proposed explanation implied that surface deformation was mostly accommodated by diffuse inelastic strain within compliant fault damage zones (e.g. Antoine et al., 2021). A similar behavior would be forecasted for the compliant HSDZ in relation to the propagation of large mainshocks (M≤7; Galli et al., 2022) along the VCFZ and the Campo Imperatore Fault System. At the same time, the shallow-velocity structure of the VCFZ, characterized by large compliant bodies in sharp contact with relatively stiffer ones (i.e. fault core, LSDZ and the host rock) is expected to influence the up-dip propagation of earthquake ruptures as suggested by numerical simulations of dynamic ruptures (e.g. Huang et al., 2014, 2016). In particular, the development of fault zone waves in low-velocity rock layers with sharp boundaries was shown to control the nature and speed of the rupture front and possibly generate new rupture pulses, with significant increase of inelastic off-fault deformation and possible ground motion amplification (Ma and Beroza, 2008; Huang et al., 2016). In relation to the latter point, a previous ambient noise seismic survey conducted across the VCFZ by Pischiutta et al. (2017) clearly highlighted a directional NNE-SSW amplification perpendicular to the average strike of the fault-fracture network and a dominant amplified frequency band of 0.5–2Hz. This polarization was related to the occurrence of a compliance anisotropy within the VCFZ, and might be partly associated to the thick low-velocity damage zone dominating its shallow structure (Fig.1). The occurrence of intensely fractured and compliant damage zones would not only determine possible ground motion amplifications with significant consequences for seismic hazard but is also crucial in controlling the near-surface mechanical stability and hydraulics of fault-controlled mountain slopes affected by landslides and deep-seated gravitational deformations (e.g. Del Rio et al., 2021; Moro et al., 2007).

## 6. Conclusions

The near-surface velocity structure of the Vado di Corno seismogenic fault zone (Central Apennines, Italy) was quantified at different length scales, from laboratory measurements of ultrasonic velocities (rock samples of few centimeters in size, 1 MHz source) to high-resolution seismic refraction tomography (spatial resolution of a few meters). Previous studies mapped in detail the fault zone footwall block and recognized distinct fault-bounded structural units: (i) fault core cataclastic units (CU1 and CU2); (ii) breccia unit (BU), (iii) high-strain damage zone (HSDZ), (iv) low-strain damage zone (LSDZ; Demurtas et al., 2016; Fondriest et al., 2020). These structural units were systematically sampled across the VCFZ and small oriented rock specimens were prepared to assess directional P- and S-wave ultrasonic velocities, and their relation with porosity and microstructures. At the length scale of few centimeters, the fault core cataclastic units were significantly "slower" (Vp 4-5 $kms^{-1}$ and Vs 2.2-3 $kms^{-1}$) compared to the damage zone units (5-6.5 $kms^{-1}$ and Vs 3-3.6 $kms^{-1}$). A general negative correlation between ultrasonic velocities and porosity was measured, with some dispersion in the fault core units related to the degree of textural maturity and pore space sealing by calcite. The occurrence of low-velocity outliers in the damage zone (Vp ~ 4.5 $kms^{-1}$ and Vs ~ 2.7 $kms^{-1}$) was instead related to the presence of microfracture networks with local cataclasis and partial sealing by calcite.

A P-wave high-resolution seismic tomography was performed in the field along a profile (~ 90 m) across the VCFZ. The tomographic results clearly illuminated fault-bounded rock bodies with distinct velocities and characterized by geometries (i.e. synthetic and antithetic extensional faults) which well compared with those deduced from the structural mapping. At the larger length scale investigated by the active seismic survey, relatively "fast" fault core units (CU1-CU2, Vp ≤ 3.0 $kms^{-1}$) and low-strain damage zone (LSDZ, Vp = 2.5-3.4 $kms^{-1}$) contrasted with a "slow" high-strain damage zone (HSDZ, Vp < 1.6 $kms^{-1}$). The discrepancy between the higher seismic velocities determined in the laboratory and the lower ones derived from the seismic tomography was investigated through a simplified effective medium approach accounting for the effect of the meso-scale fractures. As a result, the different scaling in seismic velocities displayed by the VCFZ structural units was shown to be mainly controlled by their distinct fracture density and fracture size distribution in the range of length 0.04-2 m. The resulting near-surface velocity structure was characterized by a large compliant high-strain damage zone and abrupt internal velocity contrasts, which might significantly impact the fault zone mechanics during the seismic cycle and specific factors such as the distribution and magnitude of co-seismic surface deformation.
Lastly, to our knowledge, this study quantifies for the first time at high-resolution the near-surface seismic velocity structure of an exhumed carbonate-hosted seismogenic source in the Italian Central Apennines. The documentation of such a heterogeneous velocity structure represents a novel observation compared the relatively homogeneous low-velocity zones detected by larger-scale seismic surveys around faults worldwide.

**Acknowledgements**
Funding was provided by the European Research Council Consolidator Grant Project NOFEAR 614705 (PI: Giulio Di Toro) and by the Marie Skłodowska-Curie Individual Fellowship DAMAGE 839880 (PI: Michele Fondriest). Michele Fondriest is funded by NextGenerationEU, 2021 STARS Grants@Unipd programme and supported by the 'The Geosciences for Sustainable Development' project (Budget Ministero dell'Università e della Ricerca–Dipartimenti di Eccellenza 2023–2027 C93C23002690001). Michele Fondriest thanks: Steve Boon and Jim Davy for technical support in the UCL laboratories; Leonardo Tauro for thin sections preparation; Nicola Michelon for thin section scans; Fabio Villani for help in deploying one seismic line; Matteo Demurtas for help in collecting rock blocks within the Vado di Corno fault zone; Nicolas Brantut for tips on ultrasonic S-wave first-arrival picking. Frans Aben, Francois Passelegue, Christophe Voisin, Eric Larose, Giulio Di Toro and Ashley Stanton-Yonge are thanked for fruitful discussions.

**Open Research**
The data supporting this work (sample locations, sample photographs, ultrasonic waveforms, thin section scans, mesoscale fracture datasets, active seismic survey data) are available in a public repository (Fondriest et al., 2025).